\documentclass[12pt,a4paper]{article}
\usepackage{graphicx}
\usepackage{cite}
\usepackage{amsmath}
\usepackage{myart}
\usepackage{amsfonts}
\usepackage{amssymb}
\begin{document}
\hfill PTA/07-02

\bigskip\ 

\hfill In memory of Yasha Belavin

\bigskip

\begin{center}
{\Large \textbf{Bootstrap in Supersymmetric Liouville Field Theory.}}

{\Large \textbf{I. NS Sector}}

\vspace{1.0cm}

{\large A.Belavin}

\vspace{0.2cm}

L.D.Landau Institute for Theoretical Physics RAS,

142432 Chernogolovka, Russia

\vspace{0.2cm}

{\large V.Belavin}\footnote{Institute for Theoretical and Experimental Physics
(ITEP) B. Cheremushkinskaya 25, 117259 Moscow, Russia}

\vspace{0.2cm}

International School for Advanced Studies (SISSA)

Via Beirut 2-4, 34014 Trieste, Italy

INFN, Sezione di Trieste

\vspace{0.3cm}

{\large A.Neveu}

\vspace{0.2cm}

{\large \ }and

\vspace{0.2cm}

{\large Al.Zamolodchikov}\footnote{On leave of absence from: Service de
Physique Th\'eorique, CNRS - URA 2306, C.E.A. - Saclay, F-91191,
Gif-sur-Yvette, France}

\vspace{0.2cm}

Laboratoire de Physique Th\'eorique et Astroparticules, UMR-5207 CNRS-UM2,

Universit\'e Montpellier II, Pl.E.Bataillon, 34095 Montpellier, France
\end{center}

\vspace{1.0cm}

\textbf{Abstract}

A four point function of basic Neveu-Schwarz exponential fields is constructed
in the $N=1$ supersymmetric Liouville field theory. Although the basic NS
structure constants were known previously, we present a new derivation, based
on a singular vector decoupling in the NS sector. This allows to stay
completely inside the NS sector of the space of states, without referencing to
the Ramond fields. The four-point construction involves also the NS blocks,
for which we suggest a new recursion representation, the so-called elliptic
one. The bootstrap conditions for this four point correlation function are
verified numerically for different values of the parameters.

\section{The $N=1$ super Liouville}

Construction of the super Liouville field theory (SLFT) is motivated by the
following action, which appeared in the non-critical superstring theory in
1981 \cite{Polyakov}
\begin{equation}
\mathcal{L}_{\text{SLFT}}=\frac1{8\pi}\left(  \partial_{a}\phi\right)
^{2}+\frac1{2\pi}\left(  \psi\bar\partial\psi+\bar\psi\partial\bar\psi\right)
+2i\mu b^{2}\bar\psi\psi e^{b\phi}+2\pi b^{2}\mu^{2}e^{2b\phi}\label{SL}%
\end{equation}
where $b$ is the standard ``quantum'' parameter related through the
``background charge'' $Q=b^{-1}+b$ to the central charge
\begin{equation}
\widehat{c}=1+2Q^{2}\label{cQ}%
\end{equation}
of the superconformal algebra generated by the supercurrent $S(z)$, $\bar
S(\bar z)$ and the stress tensor $T(z)$, $\bar T(\bar z)$. Traditionally the
scale parameter $\mu$ is called the (super) cosmological constant.

Let us recall some details about SLFT necessary for the forthcoming discussion
(see e.g., \cite{ZP, MSS, DHK, AG, Rubik}). The space of fields splits into
the so called Neveu-Schwarz\cite{NS} (NS) and Ramond\cite{R} (R) sectors, the
supercurrent components $(S$,$\bar S)$ being respectively single or double
valued near the point where the corresponding operator is located. Apparently
the first, NS sector, is closed under the operator product expansions (OPE).
It is completely consistent to consider it separately. This is what we're
going to do in the present publication, meaning to include the R sector in the future.

Respectively, the NS fields belong to highest weight representation of the NS
superconformal algebra
\begin{align}
\{G_{k},G_{l}\} &  =2L_{k+l}+\frac{\widehat{c}}2\left(  k^{2}-\frac14\right)
\delta_{k+l}\nonumber\\
\lbrack L_{n},G_{k}] &  =\left(  \frac n2-k\right)  G_{n+k}\label{SVir}\\
\lbrack L_{m},L_{n}] &  =(m-n)L_{m+n}+\frac{\widehat{c}}8(m^{3}-m)\delta
_{m+n}\nonumber
\end{align}
where the subscripts $m,n$ are integer and $k,l$ are half-integer. In fact
there are two copies of algebra (\ref{SVir}), the ``right'' one $SVir$
generated by $S(z)$ and $T(z),$ and the ``left'' $\overline{SVir}$ constructed
from the left-moving components $\bar S(\bar z)$ and $\bar T(\bar z)$. The
space is classified in the highest weight representations of $SVir\otimes
\overline{SVir}$. The basic NS fields are the scalar primary fields $V_{a}(x)$
corresponding to the highest weight vectors
\begin{align}
L_{n}V_{a}  & =0\,;\;\;\,\,\bar L_{n}V_{a}=0\,;\;\text{for}\;n>0\nonumber\\
\;\;G_{k}V_{a}  & =0\,;\;\;\;\bar G_{k}V_{a}=0\,;\;\text{for}%
\;k>0\label{highest}\\
L_{0}V_{a}  & =\bar L_{0}V_{a}=\Delta_{a}V_{a}\nonumber
\end{align}
where
\begin{equation}
\Delta_{a}=\frac{a(Q-a)}2\label{Da}%
\end{equation}
and $a$ is a (complex) continuous parameter. It is sometimes instructive to
think of these basic operators as of the properly regularized exponentials
$V_{a}=\exp(a\phi)$ of the fundamental bosonic field entering the Lagrangian
(\ref{SL}). This is particularly useful in the region of the configuration
space where $\phi\rightarrow-\infty$. Here one can neglect the interaction
terms in the action (\ref{SL}), the fields $\phi$ and $(\psi,\bar\psi)$ behave
as a free boson and a free Majorana fermion and the exponential expression can
be given an exact sense.

All other NS fields are the $SVir\otimes\overline{SVir}$ descendents of these
basic ones. It will prove convenient to distinguish the descendents of integer
and half-integer level, for which we reserve (somewhat loosely) the terms
``even'' descendents, often marked by the index ``e'', and the ``odd'' ones,
referred to as ``o''. It will be also useful to introduce special notations
for the components of the multiplets under the standard super Poincare
algebra, a subalgebra of $SVir\otimes\overline{SVir}$ generated by $G_{-1/2}$,
$\bar G_{-1/2}$ and $L_{0}-\bar L_{0}$. From this point of view $V_{a}$ is the
``bottom'' component of the supermultiplet, which includes also
\begin{align}
\Lambda_{a}  & =G_{-1/2}V_{a}=-ia\psi e^{a\,\phi}\nonumber\\
\bar\Lambda_{a}  & =\bar G_{-1/2}V_{a}=-ia\bar\psi e^{a\,\phi}\label{W}\\
W_{a}  & =G_{-1/2}\bar G_{-1/2}V_{a}=a^{2}\bar\psi\psi e^{a\,\phi}-2i\pi\mu
abe^{(a+b)\phi}\nonumber
\end{align}
Here we partially borrow apt notations from ref.\cite{Fukuda} and also give a
``free field'' interpretations of the corresponding components. The basic Ward
identities are
\begin{align}
T(z)V_{a}(0) &  =\frac{\Delta_{a}}{z^{2}}V_{a}(0)+\frac1z\partial
V_{a}(0)+\text{reg}\nonumber\\
T(z)\Lambda_{a}(0) &  =\frac{\Delta_{a}+1/2}{z^{2}}\Lambda_{a}(0)+\frac
1z\partial\Lambda_{a}(0)+\text{reg}\label{WI}\\
S(z)V_{a}(0) &  =\frac1z\Lambda_{a}(0)+\text{reg}\nonumber\\
S(z)\Lambda_{a}(0) &  =\frac{2\Delta_{a}}{z^{2}}V_{a}(0)+\frac1z\partial
V_{a}(0)+\text{reg}\nonumber
\end{align}
We explicitly present the holomorphic relations and quote them for $V_{a}$ and
$\Lambda_{a}$ only. The ``right'' superconformal properties of the doublet
$\bar\Lambda_{a},W_{a}$ are the same as of $V_{a},\Lambda_{a}$ and the
``left'' ones of $V_{a},\bar\Lambda_{a}$ and $\Lambda_{a},W_{a}$ are similar
to (\ref{WI}) with obvious modifications caused by the anticommutativity of
the right and left ``fermionic'' generators $G_{k}$ and $\bar G_{k}$.

Local properties of SLFT in the NS sector are encoded in the basic operator
product expansion (here and below for the sake of brevity we denote
$\Delta=\Delta_{Q/2+iP}$ and $\Delta_{i}=\Delta_{a_{i}}$, wherever it cannot
cause any misunderstanding)
\begin{align}
\  & V_{a_{1}}(x)V_{a_{2}}(0)=\label{VV}\\
& \ \ \ \ \int\frac{dP}{4\pi}\left(  x\bar x\right)  ^{\Delta-\Delta
_{1}-\Delta_{2}}\left(  \mathbb{C}_{a_{1},a_{2}}^{Q/2+iP}\left[
V_{Q/2+iP}(0)\right]  _{\text{ee}}+\mathbb{\tilde C}_{a_{1},a_{2}}%
^{Q/2+iP}\left[  V_{Q/2+iP}(0)\right]  _{\text{oo}}\right) \nonumber
\end{align}
This OPE is continuous and involves integration over the ``momentum'' $P$.
Precisely as in the bosonic Liouville field theory \cite{LFT} the integration
contour is basically along the real axis, but should be deformed sometimes
under analytic continuation in the parameters $a_{1}$ and $a_{2}$. It is a
good idea to make such deformations explicit collecting the result in the form
of the so called discrete terms \cite{LFT}. In (\ref{VV}) $\left[
V_{p}\right]  $ denotes the contribution of the primary field $V_{p}$ and its
superconformal descendents to the operator product expansion. Unlike the
standard conformal symmetry, not all these contributions are prescribed
unambiguously by the superconformal invariance, the even and the odd ones
entering independently. From here come two different structure constants
$\mathbb{C}_{a_{1}a_{2}}^{p}$ and $\mathbb{\tilde C}_{a_{1}a_{2}}^{p}$ in the
OPE (\ref{VV}), while $\left[  V_{a}\right]  _{\text{ee}}$ and $\left[
V_{a}\right]  _{\text{oo}}$ denote respectively the collections of
``even-even''\footnote{This means even in the left and even in the right
sector. Terms ``odd-odd'', ``even-odd'' etc. have similar sense.} and
``odd-odd''descendents. As usual \cite{BPZ} both towers of descendents enjoy
the factorization in the product of holomorphic and antiholomorphic
``chains''
\begin{align}
\left[  V_{a}(0)\right]  _{\text{ee}}  & =\mathcal{C}_{\text{e}}^{\Delta
_{1},\Delta_{2}}(\Delta_{a},x)\;\overline{\mathcal{C}}_{\text{e}}^{\Delta
_{1},\Delta_{2}}(\Delta_{a},\bar x)\;V_{a}(0)\label{eodesc}\\
\left[  V_{a}(0)\right]  _{\text{oo}}  & =\mathcal{C}_{\text{o}}^{\Delta
_{1},\Delta_{2}}(\Delta_{a},x)\;\overline{\mathcal{C}}_{\text{o}}^{\Delta
_{1},\Delta_{2}}(\Delta_{a},\bar x)\;V_{a}(0)\nonumber
\end{align}
where each of the ``chain operators''
\begin{align}
\mathcal{C}_{\text{e}}^{\Delta_{1},\Delta_{2}}(\Delta,x)  & =1+x\frac
{\Delta+\Delta_{1}-\Delta_{2}}{2\Delta}L_{-1}+O(x^{2})\label{Chains}\\
\mathcal{C}_{\text{o}}^{\Delta_{1},\Delta_{2}}(\Delta,x)  & =\frac{x^{1/2}%
}{2\Delta}G_{-1/2}+O\left(  x^{3/2}\right) \nonumber
\end{align}
(and the same for $\overline{\mathcal{C}}_{\text{e}}$ and $\overline
{\mathcal{C}}_{\text{o}}$ with the ``right'' $SVir$ operators $G_{k}$ and
$L_{n}$ replaced by the ``left'' ones $\bar G_{k}$ and $\bar L_{n}$) is
determined uniquely by superconformal symmetry once the normalization of the
first term is fixed, e.g., as in (\ref{Chains}), essentially in the same way
as it occurs in the usual CFT case.

The basic NS structure constants $\mathbb{C}_{a_{1}a_{2}}^{Q/2-iP}$ and
$\mathbb{\tilde C}_{a_{1},a_{2}}^{Q/2-iP}$ in (\ref{VV}) have been evaluated
through the bootstrap technique quite a while ago in refs. \cite{Rubik,
Marian}. Here we quote their result in terms of the three-point functions
\begin{align}
\left\langle V_{a_{1}}(x_{1})V_{a_{2}}(x_{2})V_{a_{3}}(x_{3})\right\rangle  &
=\frac{C_{a_{1},a_{2},a_{3}}}{(x_{12}\bar x_{12})^{\Delta_{1+2-3}}(x_{23}\bar
x_{23})^{\Delta_{2+3-1}}(x_{31}\bar x_{31})^{\Delta_{3+1-2}}}\label{threep}\\
\left\langle W_{a_{1}}(x_{1})W_{a_{2}}(x_{2})W_{a_{3}}(x_{3})\right\rangle  &
=\frac{(1/2-\Delta_{1}-\Delta_{2}-\Delta_{3})^{2}\tilde C_{a_{1},a_{2},a_{3}}%
}{(x_{12}\bar x_{12})^{\Delta_{1+2-3}+1/2}(x_{23}\bar x_{23})^{\Delta
_{2+3-1}+1/2}(x_{31}\bar x_{31})^{\Delta_{3+1-2}+1/2}}\nonumber
\end{align}
Here and henceforth we denote $x_{ij}=x_{i}-x_{j}$ and also use the
abbreviations like $\Delta_{1+2-3}=\Delta_{1}+\Delta_{2}-\Delta_{3}$ etc. All
the other three point functions of different supermultiplet components are
expressed through these via the superprojective invariance $SL(2|1)\otimes
\overline{SL(2|1)}$. For example
\begin{align}
\left\langle W_{a_{1}}(x_{1})W_{a_{2}}(x_{2})V_{a_{3}}(x_{3})\right\rangle  &
=\frac{(\Delta_{1}+\Delta_{2}-\Delta_{3})^{2}C_{a_{1},a_{2},a_{3}}}%
{(x_{12}\bar x_{12})^{\Delta_{1+2-3}+1}(x_{23}\bar x_{23})^{\Delta_{2+3-1}%
}(x_{31}\bar x_{31})^{\Delta_{3+1-2}}}\label{oc}\\
\left\langle W_{a_{1}}(x_{1})V_{a_{2}}(x_{2})V_{a_{3}}(x_{3})\right\rangle  &
=\frac{\tilde C_{a_{1},a_{2},a_{3}}}{(x_{12}\bar x_{12})^{\Delta_{1+2-3}%
+1/2}(x_{23}\bar x_{23})^{\Delta_{2+3-1}-1/2}(x_{31}\bar x_{31})^{\Delta
_{3+1-2}+1/2}}\nonumber
\end{align}
The superconformal symmetry allows to express all three point functions with
different components (\ref{W}) of the supermultiplets through the two
constants $C_{a_{1}a_{2}a_{3}}$ and $\tilde C_{a_{1}a_{2}a_{3}}$\footnote{A
derivation based on the superprojective Ward identities, which does not
exploit a superfield formalism, is presented in Appendix A.}. Both constants
$C_{a_{1}a_{2}a_{3}}$ and $\tilde C_{a_{1}a_{2}a_{3}}$ are symmetric in all
their arguments and related to the structure constants as (normalizations
chosen in (\ref{Chains}) are important in these relations)
\begin{align}
\mathbb{C}_{a_{1}a_{2}}^{Q-p}=C_{a_{1}a_{2}p}\,;\;\;\;\;\;\mathbb{\tilde
C}_{a_{1}a_{2}}^{Q-p}=-\tilde C_{a_{1}a_{2}p}\label{CCC}%
\end{align}
Following \cite{Rubik, Marian} they have the following explicit form ($a$
stands here for $a_{1}+a_{2}+a_{3}$)
\begin{align}
C_{a_{1}a_{2}a_{3}}  & =\left(  \pi\mu\gamma\left(  \frac{Qb}2\right)
b^{1-b^{2}}\right)  ^{(Q-a)/b}\frac{\Upsilon_{\text{NS}}^{\prime}%
(0)\Upsilon_{\text{NS}}(2a_{1})\Upsilon_{\text{NS}}(2a_{2})\Upsilon
_{\text{NS}}(2a_{3})}{\Upsilon_{\text{NS}}(a-Q)\Upsilon_{\text{NS}}%
(a_{1+2-3})\Upsilon_{\text{NS}}(a_{2+3-1})\Upsilon_{\text{NS}}(a_{3+1-2}%
)}\label{C3}\\
\tilde C_{a_{1}a_{2}a_{3}}  & =\left(  \pi\mu\gamma\left(  \frac{Qb}2\right)
b^{1-b^{2}}\right)  ^{(Q-a)/b}\frac{2i\Upsilon_{\text{NS}}^{\prime}%
(0)\Upsilon_{\text{NS}}(2a_{1})\Upsilon_{\text{NS}}(2a_{2})\Upsilon
_{\text{NS}}(2a_{3})}{\Upsilon_{\text{R}}(a-Q)\Upsilon_{\text{R}}%
(a_{1+2-3})\Upsilon_{\text{R}}(a_{2+3-1})\Upsilon_{\text{R}}(a_{3+1-2}%
)}\nonumber
\end{align}
where we make use of convenient notations from ref.\cite{Fukuda} for the
special functions
\begin{align}
\Upsilon_{\text{NS}}(x)  & =\Upsilon_{b}\left(  \frac x2\right)  \Upsilon
_{b}\left(  \frac{x+Q}2\right) \label{YNSR}\\
\Upsilon_{\text{R}}(x)  & =\Upsilon_{b}\left(  \frac{x+b}2\right)
\Upsilon_{b}\left(  \frac{x+b^{-1}}2\right) \nonumber
\end{align}
expressed in terms of the standard in the Liouville field theory ``upsilon''
function $\Upsilon_{b}$ (see \cite{DO, LFT}). For us the following functional
relations are important
\begin{align}
\Upsilon_{\text{NS}}(x+b)  & =b^{-bx}\gamma\left(  \frac12+\frac{bx}2\right)
\Upsilon_{\text{R}}(x)\label{Yshift}\\
\Upsilon_{\text{R}}(x+b)  & =b^{1-bx}\gamma\left(  \frac{bx}2\right)
\Upsilon_{\text{NS}}(x)\nonumber
\end{align}
and the same with $b$ replaced by $b^{-1}$. Finally
\begin{equation}
\Upsilon_{\text{NS}}^{\prime}(0)=\frac12\Upsilon_{b}\left(  \frac b2\right)
\Upsilon_{b}\left(  \frac1{2b}\right) \label{Y0}%
\end{equation}

All these expressions correspond to the ``natural'' normalization of the
exponential fields, where the two-point function reads
\begin{equation}
\left\langle V_{a}(x)V_{a}(0)\right\rangle =\frac{D_{\text{NS}}(a)}{(x\bar
x)^{2\Delta_{a}}}\label{C2}%
\end{equation}
with
\begin{equation}
D_{\text{NS}}(a)=\left(  \pi\mu\gamma\left(  \frac{bQ}2\right)  \right)
^{(Q-2a)b^{-1}}\frac{b^{2}\gamma\left(  ba-\frac12-\frac{b^{2}}2\right)
}{\gamma\left(  \frac12+\frac{b^{-2}}2-ab^{-1}\right)  }\label{D}%
\end{equation}
As usual, in the natural normalization the fields $V_{a}$ and $V_{Q-a}$ are
identified through the following ``reflection relations''
\begin{equation}
V_{a}=D_{\text{NS}}(a)V_{Q-a}\label{reflection}%
\end{equation}

In refs.\cite{Rubik, Marian} (see also \cite{Fukuda}) expressions (\ref{C3})
were derived on the basis of the singular vector decoupling in the singular
Ramond representation. This approach has many advantages, being the simplest
known and giving simultaneously the OPE structure constants in the whole space
of fields, Neveu-Schwarz and Ramond. In this note we find it instructive,
however, to rederive (\ref{C3}) with the help of pure NS bootstrap, requiring
the NS null-vector decoupling. Albeit technically more complicated, this
derivation allows to stay completely inside the NS sector of SLFT and provides
a good exercise in classical analysis. This program is described in the
following section.

Once the structure constants are determined, expression (\ref{VV}) gives
directly the integral representation for the four point function of four
``bottom'' NS primary fields
\begin{align}
\  & \left\langle V_{a_{1}}(x_{1})V_{a_{2}}(x_{2})V_{a_{3}}(x_{3})V_{a_{4}%
}(x_{4})\right\rangle =\label{fourp}\\
& \ \left(  x_{41}\bar x_{41}\right)  ^{-2\Delta_{1}}\left(  x_{24}\bar
x_{24}\right)  ^{\Delta_{1+3-2-4}}\left(  x_{34}\bar x_{34}\right)
^{\Delta_{1+2-3-4}}\left(  x_{23}\bar x_{23}\right)  ^{\Delta_{4-1-2-3}%
}G\left(  \left.
\begin{array}
[c]{cc}%
a_{1} & a_{3}\\
a_{2} & a_{4}%
\end{array}
\right|  x,\bar x\right) \nonumber
\end{align}
where
\begin{equation}
x=\frac{x_{12}x_{34}}{x_{23}x_{41}}\label{x}%
\end{equation}
and the ``reduced'' four point function admits the following ``s-channel''
representation
\begin{align}
G\left(  \left.
\begin{array}
[c]{cc}%
a_{1} & a_{3}\\
a_{2} & a_{4}%
\end{array}
\right|  x,\bar x\right)   & =\int\frac{dP}{4\pi}\mathbb{C}_{a_{1}a_{2}%
}^{Q/2-iP}\mathbb{C}_{a_{3}a_{4}}^{Q/2+iP}\mathcal{F}_{\text{e}}\left(
\begin{array}
[c]{cc}%
\Delta_{1} & \Delta_{3}\\
\Delta_{2} & \Delta_{4}%
\end{array}
\left|  \Delta\right|  x\right)  \mathcal{F}_{\text{e}}\left(
\begin{array}
[c]{cc}%
\Delta_{1} & \Delta_{3}\\
\Delta_{2} & \Delta_{4}%
\end{array}
\left|  \Delta\right|  \bar x\right) \label{schan}\\
& \ -\int\frac{dP}{4\pi}\mathbb{\tilde C}_{a_{1}a_{2}}^{Q/2-iP}\mathbb{\tilde
C}_{a_{3}a_{4}}^{Q/2+iP}\mathcal{F}_{\text{o}}\left(
\begin{array}
[c]{cc}%
\Delta_{1} & \Delta_{3}\\
\Delta_{2} & \Delta_{4}%
\end{array}
\left|  \Delta\right|  x\right)  \mathcal{F}_{\text{o}}\left(
\begin{array}
[c]{cc}%
\Delta_{1} & \Delta_{3}\\
\Delta_{2} & \Delta_{4}%
\end{array}
\left|  \Delta\right|  \bar x\right) \nonumber
\end{align}
For simplicity we omit possible discrete terms. Superconformal NS blocks
$\mathcal{F}_{\text{e}}$ and $\mathcal{F}_{\text{o}}$ effectively sum up
respectively ``even'' and ``odd'' descendants of the primary NS field of
dimension $\Delta=Q^{2}/8+P^{2}/2$. Notice the minus sign in front of the
second term in (\ref{schan}). It is due to the anticommutativity of the
``right'' and ``left'' odd chain operators in (\ref{eodesc})\footnote{This
convention is consistent with the minus sign in the second relation in
eq.(\ref{CCC}). The pure imaginary structure constant $\mathbb{C}_{a_{1}a_{2}%
}^{p}$ in (\ref{C3}) assures positive sign before the net contribution of the
odd levels (see below).}. The blocks are constructed unambiguously on the
basis of superconformal invariance. The problem of their evaluation has been
addressed recently in refs.\cite{vbelavin, leshek} and will be reconsidered
below in sections 3, 4 and 5. First we recapitulate the so-called recursive
``c-representation'' \cite{cblock}, which has been developed in
\cite{vbelavin, leshek} and allows to effectively reconstruct the blocks
iteratively as a series in the parameter $x$. Better convergent is the
so-called ``delta'', or ``elliptic'' representation, whose recursions give a
much faster convergent series in the elliptic parameter $q$. Such
representation has been constructed in \cite{dblock} for the ordinary
conformal algebra. Presently in section 5 we suggest a generalization of the
elliptic representation for the $N=1$ superconformal case. In this paper we
restrict ourselves only to the blocks with all four external fields ``bottom''
supermultiplet components (and those related to this case by the
superprojective symmetry).

An important property of a consistent euclidean quantum field theory is the
associativity of the algebra of operator product expansions. This property
ensures the correlation functions to be single-valued over the euclidean slice
of the complex space-time. It is commonly believed to give an euclidean
interpretation of the standard Minkowskian locality. In particular, the
four-point function is single-valued (or, enjoys the crossing symmetry) if the
following two identities hold
\begin{align}
G\left(  \left.
\begin{array}
[c]{cc}%
a_{1} & a_{3}\\
a_{2} & a_{4}%
\end{array}
\right|  x,\bar x\right)   & =[(1-x)(1-\bar x)]^{-2\Delta_{1}}G\left(
\left.
\begin{array}
[c]{cc}%
a_{1} & a_{4}\\
a_{2} & a_{3}%
\end{array}
\right|  \frac x{x-1},\frac{\bar x}{\bar x-1}\right) \nonumber\\
G\left(  \left.
\begin{array}
[c]{cc}%
a_{1} & a_{3}\\
a_{2} & a_{4}%
\end{array}
\right|  x,\bar x\right)   & =G\left(  \left.
\begin{array}
[c]{cc}%
a_{1} & a_{2}\\
a_{3} & a_{4}%
\end{array}
\right|  1-x,1-\bar x\right) \label{cross}%
\end{align}
In the form (\ref{schan}) the first relation is a trivial consequence of the
transformation properties of the blocks. The second, however, becomes a
non-trivial relation for the structure constants. Its closed formulation in
terms of the structure constants requires explicit knowledge of the fusion (or
crossing) matrix for a general superconformal block, which is not currently
available (see \cite{Ponsot} for an explicit construction in the bosonic
case). Under these circumstances, we find it of value to verify this relation
numerically, taking the explicit form of the structure constants and using the
fast convergent elliptic representation for the superconformal blocks. This we
perform in section 6 and find a reasonable numerical support for the relations
(\ref{cross}) for certain randomly chosen values of the parameters.

Section 7 is devoted to the discussion and outlook.

\section{Singular vector bootstrap in NS sector}

At certain values of the parameter $a$ the SLFT operator $V_{a}$ is a highest
vector of a singular representation of $SVir$. In the NS sector, which we only
consider in this paper, the simplest singular representation has dimension
$\Delta_{1,3}=-1/2-b^{2}$ with the primary field $V_{-b}$ as the highest
weight vector. The singular vector appearing at the level $3/2$ reads
\begin{equation}
\left(  G_{-1/2}^{3}+b^{2}G_{-3/2}\right)  V_{-b}\label{SV13}%
\end{equation}
There is also a singular vector of the same form in the left $\overline{SVir}
$ sector. Vanishing of all singular vectors in the singular representations
can be taken as the basic dynamical principle of SLFT, precisely like in the
ordinary bosonic LFT. In particular, setting (\ref{SV13}) and its ``left''
counterpart to zero leads to non-trivial equations for the correlation
functions. It is easy to see, e.g., considering the three point function with
the field $V_{-b}$, that such ``decoupling equation'' restricts the form of
the operator product expansion of a product of this degenerate field and
arbitrary primary $V_{a}$ to the following ``discrete'' form
\begin{align}
\  & V_{-b}V_{a}=\label{OPE13}\\
& \ \ \ \ (x\bar x)^{ab}C_{-}(a)\left[  V_{a-b}\right]  _{\text{ee}}+(x\bar
x)^{1/2+b^{2}}\tilde C_{0}(a)\left[  V_{a}\right]  _{\text{oo}}+C_{+}(a)(x\bar
x)^{1-ba+b^{2}}\left[  V_{a+b}\right]  _{\text{ee}}\nonumber
\end{align}
where $C_{-}(a)$, $\tilde C_{0}(a)$ and $C_{+}(a)$ are ``special'' (unlike the
``generic'' ones in the OPE (\ref{VV})), or ``discrete'' structure constants.
It is instructive to understand how the general OPE (\ref{VV}) with the
structure constants (\ref{C3}) results in the discrete expansion (\ref{OPE13})
if one of the parameters $a_{1}$ or $a_{2}$ is set to the singular value $-b$.
This calculation is performed in the Appendix B.

Below in this section we will use the bootstrap conditions together with the
``decoupling principle'' to derive unambiguously the generic NS structure
constants (\ref{C3}). In particular, the values of the special structure
constants in (\ref{OPE13}) are recovered uniquely up to an overall scale. It
is instructive, however, to give a ``perturbative'' derivation, similar to
that presented e.g., in \cite{FZZ}, because firstly it is simpler and more
transparent, and secondly it gives a heuristic link with the SLFT Lagrangian
(\ref{SL}), in particular relates the scale parameter to the cosmological
constant $\mu$. The idea is simply a version of the ``screening'' calculus,
invented by B.Feigin and D.Fuchs (and further developed by V.Fateev and
V.Dotsenko \cite{DF}) where the role of the screening operator is played by
the interaction term $2i\mu W_{b}$ in (\ref{SL}). E.g., neglecting formally
this term, one considers $\phi$ and $\psi$ as free fields, so that the
exponentials $V_{-b}$ and $V_{a}$ enjoy the free field fusion to $V_{a-b}$
with
\begin{equation}
C_{-}(a)=1\label{Cm}%
\end{equation}
The next term with $\left[  V_{a}\right]  _{\text{oo}}=(2\Delta_{a}%
)^{-2}(x\bar x)^{1/2}W_{a}+\ldots$ requires one insertion of the perturbation
$2i\mu b^{2}\int\bar\psi\psi e^{b\phi}d^{2}x$. Thus
\begin{align}
\tilde C_{0}(a)  & =-2i\mu b^{2}(Q-a)^{2}\int\left\langle \bar\psi\psi
e^{b\phi}(y)V_{a}(0)V_{-b}(1)\bar\psi\psi e^{(Q-a)\phi}\left(  \infty\right)
\right\rangle \nonumber\\
\  & =\frac{2\pi i\mu\gamma(ab-b^{2})}{\gamma(-b^{2})\gamma(ab)}\label{C0}%
\end{align}
Finally, we need to make two perturbative insertions in order to create the
last term $\left[  V_{a+b}\right]  _{\text{ee}}=V_{a+b}+\ldots$ in the OPE
(\ref{OPE13}). This results in
\begin{align}
C_{+}(a)  & =\frac{\left(  -2i\mu b^{2}\right)  ^{2}}2\int\left\langle
\bar\psi\psi e^{b\phi}(y_{1})\bar\psi\psi e^{b\phi}(y_{2})V_{a}(0)V_{-b}%
(1)V_{Q-a-b}(\infty)\right\rangle \label{Cp}\\
\  & =2\mu^{2}b^{4}\int[(y_{1}-y_{2})(\bar y_{1}-\bar y_{2})]^{-b^{2}-1}%
\prod_{i=1}^{2}(y_{i}\bar y_{i})^{-ab}\left[  (1-y_{i})(1-\bar y_{i})\right]
^{b^{2}}d^{2}y_{i}\nonumber\\
\  & =\pi^{2}\mu^{2}b^{4}\gamma^{2}\left(  \frac12+\frac{b^{2}}2\right)
\gamma\left(  -\frac12-\frac{b^{2}}2+ab\right)  \gamma\left(  \frac
12-\frac{b^{2}}2-ab\right) \nonumber
\end{align}
In the last calculation we have used the following integration formula
\begin{align}
& \ \ \ \ \ \frac1{n!}\int\prod_{i=1}^{n}(z_{i}\bar z_{i})^{\mu-1}%
[(1-z_{i})(1-\bar z_{i})]^{\nu-1}d^{2}z_{i}\prod_{i>j}[(z_{i}-z_{j})(\bar
z_{i}-\bar z_{j})]^{2g}\label{DFintegral}\\
\  & =\left(  \pi\gamma(1-g)\right)  ^{n}\prod_{k=0}^{n-1}\gamma
(g+kg)\gamma(\mu+kg)\gamma(\nu+kg)\gamma(\lambda+kg)\nonumber
\end{align}
where $\mu+\nu+\lambda=1-2(n-1)g$. This formula belongs to Dotsenko and Fateev
\cite{FD2}.

It is a simple exercise in operator product expansion (see e.g. \cite{FZZ}) to
show that the two-point function (\ref{C2}) satisfies the functional relation
\begin{equation}
\frac{D_{\text{NS}}(a)}{D_{\text{NS}}(a+b)}=C_{+}(a)\label{DDC}%
\end{equation}
This is indeed the case for (\ref{D}) together with (\ref{Cp}). The ``dual''
functional relation
\begin{equation}
\frac{D_{\text{NS}}(a)}{D_{\text{NS}}(a+b^{-1})}=\pi^{2}\tilde\mu^{2}%
b^{4}\gamma^{2}\left(  \frac12+\frac{b^{-2}}2\right)  \gamma\left(
-\frac12-\frac{b^{-2}}2+ab^{-1}\right)  \gamma\left(  \frac12-\frac{b^{-2}%
}2-ab^{-1}\right) \label{DDual}%
\end{equation}
is also satisfied if the ``dual cosmological constant'' $\tilde\mu$ is related
to $\mu$ as
\begin{equation}
\pi\tilde\mu\gamma\left(  \frac12+\frac{b^{-2}}2\right)  =\left(  \pi\mu
\gamma\left(  \frac12+\frac{b^{2}}2\right)  \right)  ^{b^{2}}\label{mudual}%
\end{equation}

Now, let us turn to the four point function with one singular primary field
$V_{-b}$ and three arbitrary ones. It is natural to renumber the operators,
setting in (\ref{fourp}) $V_{a_{1}}$ to be $V_{-b}$ and denoting $V_{a_{2}}$,
$V_{a_{3}}$ and $V_{a_{4}}$ as $V_{1}$, $V_{2}$ and $V_{3}$, their dimensions
being $\Delta_{1}$, $\Delta_{2}$ and $\Delta_{3}$ respectively. Thus, in the
notations of (\ref{fourp})
\begin{equation}
g(x,\bar x)=G\left(  \left.
\begin{array}
[c]{cc}%
-b & a_{2}\\
a_{1} & a_{3}%
\end{array}
\right|  x,\bar x\right) \label{g}%
\end{equation}
It has been shown in \cite{vbelavin} that this function satisfies the
following third order linear differential equation
\begin{align}
& \ \ \ \frac1{b^{2}}g^{\prime\prime\prime}+\frac{1-2b^{2}}{b^{2}}\frac
{1-2x}{x(1-x)}g^{\prime\prime}+\left(  \frac{b^{2}+2\Delta_{1}}{x^{2}}%
+\frac{b^{2}+2\Delta_{2}}{(1-x)^{2}}+\frac{2-3b^{2}+2\Delta_{1+2-3}}%
{x(1-x)}\right)  g^{\prime}+\label{diffeq}\\
\  & +\left(  \frac{2\Delta_{2}(1+b^{2})}{(1-x)^{3}}-\frac{2\Delta_{1}%
(1+b^{2})}{x^{3}}+\frac{\Delta_{2-1}+(1-2x)(b^{4}+b^{2}(1/2-\Delta
_{1+2-3})-\Delta_{1+2})}{x^{2}(1-x)^{2}}\right)  g=0\nonumber
\end{align}
The same equation holds with respect to $\bar x$. Near $x\rightarrow0$ it has
three exponents $ba_{1}$, $1+b^{2}$ and $1-ba_{1}+b^{2}$, which correspond,
respectively, to the dimensions $\Delta_{a_{1}-b}$, $\Delta_{a_{1}}+1/2$ and
$\Delta_{a_{1}+b}$. They give rise to three s-channel blocks
\begin{align}
\mathcal{F}_{\text{e}}^{(-)}(x)  & =x^{a_{1}b}\left(  1+\ldots\right)
\nonumber\\
\mathcal{F}_{\text{o}}^{(0)}(x)  & =x^{1+b^{2}}\left(  \frac1{2\Delta_{a_{1}}%
}+\ldots\right) \label{blocks13}\\
\mathcal{F}_{\text{e}}^{(+)}(x)  & =x^{1-ba_{1}+b^{2}}\left(  1+\ldots\right)
\nonumber
\end{align}
where in the brackets stand regular series in $x$ and the normalization
correspond to the general convention of (\ref{Chains}). The correlation
function is combined as
\begin{align}
g(x,\bar x)  & =-\tilde C_{0}(a_{1})\tilde C_{a_{1},a_{2},a_{3}}%
\mathcal{F}_{\text{o}}^{(0)}(x)\mathcal{F}_{\text{o}}^{(0)}(\bar
x)\label{gFF}\\
& +C_{-}(a_{1})C_{a_{1}-b,a_{2},a_{3}}\mathcal{F}_{\text{e}}^{(-)}%
(x)\mathcal{F}_{\text{e}}^{(-)}(\bar x)+C_{+}(a_{1})C_{a_{1}+b,a_{2},a_{3}%
}\mathcal{F}_{\text{e}}^{(+)}(x)\mathcal{F}_{\text{e}}^{(+)}(\bar x)\nonumber
\end{align}

It turns out that (\ref{diffeq}) is of the type considered by Dotsenko and
Fateev in \cite{DF} and can be solved in terms of two-fold contour integrals.
Explicitly
\begin{align}
\mathcal{F}_{\text{e}}^{(-)}(x)  & =\frac{x^{a_{1}b}(1-x)^{a_{2}b}%
\Gamma\left(  -\frac12-\frac{b^{2}}2\right)  \Gamma\left(  \frac12-\frac
{b^{2}}2+ba_{1}\right)  \Gamma(ba_{1}-b^{2})F_{-}(A,B,C,g;x)}{\Gamma
(-1-b^{2})\Gamma\left(  \frac12+\frac{ba_{1+3-2}}2\right)  \Gamma\left(
\frac{ba_{1+3-2}}2-\frac{b^{2}}2\right)  \Gamma\left(  \frac12+\frac
{ba_{1+2-3}}2\right)  \Gamma\left(  \frac{ba_{1+2-3}}2-\frac{b^{2}}2\right)
}\nonumber\\
\mathcal{F}_{\text{o}}^{(0)}(x)  & =\frac{x^{1+b^{2}}(1-x)^{a_{2}b}b^{2}%
\Gamma\left(  1-ba_{1}+b^{2}\right)  \Gamma(ba_{1})F_{0}(A,B,C,g;x)}%
{\Gamma\left(  \frac12+\frac{ba_{2+3-1}}2\right)  \Gamma\left(  \frac
12+\frac{ba_{1+3-2}}2\right)  \Gamma\left(  \frac32-\frac{ba_{1+2+3}}%
2+b^{2}\right)  \Gamma\left(  \frac12+\frac{ba_{1+2-3}}2\right)  }\label{FI12}%
\end{align}
and
\begin{align}
\  & \mathcal{F}_{\text{e}}^{(+)}(x)=\label{FI3}\\
& \ \ \frac{x^{1-ba_{1}+b^{2}}(1-x)^{a_{2}b}\Gamma\left(  -\frac12-\frac
{b^{2}}2\right)  \Gamma\left(  \frac32-ba_{1}+\frac{b^{2}}2\right)
\Gamma\left(  1-ba_{1}\right)  F_{+}(A,B,C,g;x)}{\Gamma(-1-b^{2})\Gamma\left(
\frac12+\frac{ba_{2+3-1}}2\right)  \Gamma\left(  \frac{ba_{2+3-1}}%
2-\frac{b^{2}}2\right)  \Gamma\left(  \frac32-\frac{ba_{1+2+3}}2+b^{2}\right)
\Gamma\left(  1-\frac{ba_{1+2+3}}2+\frac{b^{2}}2\right)  }\nonumber
\end{align}
Functions $F_{\pm}(A,B,C,g;x)$ and $F_{0}(A,B,C,g;x)$ are regular series in $x
$ and admit the following integral representations
\begin{align}
F_{-}(x)  & =%
{\displaystyle\int\limits_{0}^{1}}
dt_{1}%
{\displaystyle\int\limits_{0}^{t_{1}}}
dt_{2}(t_{1}-t_{2})^{2g}(t_{1}t_{2})^{-A-B-C-2}[(1-t_{1})(1-t_{2}%
)]^{B}[(1-xt_{1})(1-xt_{2})]^{C}\nonumber\\
F_{+}(x)  & =%
{\displaystyle\int\limits_{0}^{1}}
dt_{1}%
{\displaystyle\int\limits_{0}^{t_{1}}}
dt_{2}(t_{1}-t_{2})^{2g}(t_{1}t_{2})^{A}[(1-t_{1})(1-t_{2})]^{C}%
[(1-xt_{1})(1-xt_{2})]^{B}\label{Fint}\\
F_{0}(x)  & =%
{\displaystyle\int\limits_{0}^{1}}
t_{1}^{-2-A-B-C-2g}(1-t_{1})^{B}(1-xt_{1})^{C}dt_{1}%
{\displaystyle\int\limits_{0}^{1}}
dt_{2}t_{2}{}^{A}[(1-xt_{2})^{B}[(1-t_{2})^{C}(1-xt_{1}t_{2})^{2g}\nonumber
\end{align}
Parameters $A$, $B$, $C$ and $g$ are related to the super Liouville ones
$a_{1} $, $a_{2}$, $a_{3}$ and $b$ as
\begin{align}
A  & =-\frac12+\frac{ba_{2+3-1}}2\,;\;\;\;\;\;\;\;B=-\frac12+\frac{ba_{1-2+3}%
}2\label{ABC}\\
C  & =\frac12-\frac{ba_{1+2+3}}2+b^{2}\,\,;\;\;\;g=-\frac12-\frac{b^{2}%
}2\nonumber
\end{align}
For the sake of completeness in Appendix C we recapitulate from \cite{DF} the
third order differential equation of Dotsenko-Fateev type, relevant integral
representations of the solutions, as well as their monodromy properties.

In particular, the combination (\ref{gFF}) is a single-valued function of the
two-dimensional coordinate $(x,\bar x)$ if
\begin{align}
\  & \frac{C_{+}(a_{1})C_{a_{1}+b,a_{2},a_{3}}}{C_{-}(a_{1})C_{a_{1}%
-b,a_{2},a_{3}}}=-\frac{\gamma\left(  ba_{1}\right)  \gamma\left(
ba_{1}-b^{2}\right)  \gamma^{2}\left(  \frac12-\frac{b^{2}}2+ba_{1}\right)
}{\left(  \frac12-ba_{1}+\frac{b^{2}}2\right)  ^{2}}\times\label{CpCp}\\
& \ \ \ \ \ \ \frac{\gamma\left(  \frac12+\frac{ba_{2+3-1}}2\right)
\gamma\left(  \frac{ba_{2+3-1}}2-\frac{b^{2}}2\right)  \gamma\left(
\frac32-\frac{ba_{1+2+3}}2+b^{2}\right)  \gamma\left(  1-\frac{ba_{1+2+3}%
}2+\frac{b^{2}}2\right)  }{\gamma\left(  \frac12+\frac{ba_{1+3-2}}2\right)
\gamma\left(  \frac{ba_{1+3-2}}2-\frac{b^{2}}2\right)  \gamma\left(
\frac12+\frac{ba_{1+2-3}}2\right)  \gamma\left(  \frac{ba_{1+2-3}}%
2-\frac{b^{2}}2\right)  }\nonumber
\end{align}
and
\begin{align}
\frac{\tilde C_{0}(a_{1})\tilde C_{a_{1},a_{2},a_{3}}}{C_{-}(a_{1}%
)C_{a_{1}-b,a_{2},a_{3}}}  & =\ \frac{-\gamma\left(  -\frac12-\frac{b^{2}%
}2\right)  \gamma\left(  \frac12-\frac{b^{2}}2+ba_{1}\right)  \gamma
^{2}(ba_{1}-b^{2})}{b^{4}\gamma\left(  -1-b^{2}\right)  \gamma\left(
ba_{1}\right)  }\times\label{C0C0}\\
& \ \ \ \ \ \ \ \ \ \frac{\gamma\left(  \frac12+\frac{ba_{2+3-1}}2\right)
\gamma\left(  \frac32-\frac{ba_{1+2+3}}2+b^{2}\right)  }{\gamma\left(
\frac{ba_{3+1-2}}2-\frac12b^{2}\right)  \gamma\left(  \frac{ba_{1+2-3}}%
2-\frac{b^{2}}2\right)  }\nonumber
\end{align}
These formulas, together with the explicit expressions (\ref{Cm}), (\ref{C0})
and (\ref{Cp}) for the special structure constants, result in the following
functional relations for the three-point functions $C_{a_{1},a_{2},a_{3}} $
and $\tilde C_{a_{1},a_{2},a_{3}}$%

\begin{align}
\  & \frac{C_{a_{1}+b,a_{2},a_{3}}}{C_{a_{1}-b,a_{2},a_{3}}}=\frac
{\gamma\left(  \frac12+\frac{b^{2}}2+a_{1}b\right)  \gamma\left(
ba_{1}\right)  \gamma\left(  ba_{1}-b^{2}\right)  \gamma\left(  \frac
12-\frac{b^{2}}2+ba_{1}\right)  }{\pi^{2}\mu^{2}b^{4}\gamma^{2}\left(
\frac12+\frac{b^{2}}2\right)  \gamma\left(  -\frac12-\frac{b^{2}}%
2+a_{1}b\right)  \gamma\left(  \frac12-\frac{b^{2}}2+a_{1}b\right)  }%
\times\label{CC}\\
& \ \ \frac{\gamma\left(  1-\frac12ba_{1+2+3}+\frac12b^{2}\right)
\gamma\left(  \frac32-\frac{ba_{1+2+3}}2+b^{2}\right)  \gamma\left(
\frac12+\frac12ba_{2+3-1}\right)  \gamma\left(  \frac12ba_{2+3-1}-\frac
12b^{2}\right)  }{\gamma\left(  \frac12+\frac12ba_{1+3-2}\right)
\gamma\left(  \frac12ba_{1+3-2}-\frac12b^{2}\right)  \gamma\left(
\frac12+\frac12ba_{1+2-3}\right)  \gamma\left(  \frac12ba_{1+2-3}-\frac
12b^{2}\right)  }\nonumber
\end{align}
and
\begin{align}
& \frac{\tilde C_{a_{1},a_{2},a_{3}}}{C_{a_{1}-b,a_{2},a_{3}}}=\label{C0C}\\
& \frac{2i\gamma\left(  \frac12-\frac{b^{2}}2+ba_{1}\right)  \gamma
(ba_{1}-b^{2})\gamma\left(  \frac12+\frac{ba_{2+3-1}}2\right)  \gamma\left(
\frac32-\frac{ba_{1+2+3}}2+b^{2}\right)  }{\pi\mu b^{4}\gamma\left(
\frac12+\frac{b^{2}}2\right)  \gamma\left(  \frac{ba_{3+1-2}}2-\frac
12b^{2}\right)  \gamma\left(  \frac{ba_{1+2-3}}2-\frac{b^{2}}2\right)
}\nonumber
\end{align}

It is verified directly through the shift relations (\ref{Yshift}) that the
structure constants (\ref{C3}) satisfy these functional relations. A standard
consideration \cite{Teschner} with the ``dual'' functional relation (i.e., the
relation with $b\rightarrow b^{-1}$ and $\mu\rightarrow\tilde\mu$) now applies
and allows to argue that, at least at real irrational values of $b^{2}$,
expressions (\ref{C3}) give the unique solution to the singular vector
decoupling equations.

\section{Analytic structure of NS block}

Here we analyze in more details the properties of the superconformal blocks
$\mathcal{F}_{\text{e}}$ and $\mathcal{F}_{\text{o}}$ which enter the integral
representation (\ref{schan}) of the four-point function. The ``chain
operators'' introduced in (\ref{eodesc}) are mostly important in this
analysis. The ``right-left'' factorization allows to concentrate on the
holomorphic part only and then combine it with the (mostly identical)
antiholomorphic one. As it has been found in \cite{leshek} and \cite{vbelavin}%
, apart from the OPE (\ref{VV}) one has to consider the similar OPE
$\Lambda_{a_{1}}(x)V_{a_{2}}(0)$. Commutation relations
\begin{align}
\lbrack G_{k},V_{a}(x)]  & =x^{k+1/2}\Lambda_{a}(x)\label{GV}\\
\{G_{k},\Lambda_{a}(x)\}  & =x^{k+1/2}\left(  \frac{2\Delta_{a}(k+1/2)}%
xV_{a}(x)+\frac\partial{\partial x}V_{a}(x)\right) \nonumber
\end{align}
follow directly from (\ref{WI}). Acting by any generator $G_{k}$ with $k>0$ on
the left hand side of (\ref{VV}) one finds
\begin{align}
\Lambda_{a_{1}}(x)V_{a_{2}}(0)  & =x^{-1/2}\int\frac{dP}{4\pi}\left(  x\bar
x\right)  ^{\Delta-\Delta_{1}-\Delta_{2}}\left(  \mathbb{C}_{a_{1},a_{2}%
}^{Q/2+iP}\mathcal{\tilde C}_{\text{o}}(\Delta_{a},x)\;\overline{\mathcal{C}%
}_{\text{e}}(\Delta_{a},\bar x)V_{Q/2+iP}(0)\right. \nonumber\\
& \ \ \ \left.  +\mathbb{\tilde C}_{a_{1},a_{2}}^{Q/2+iP}\mathcal{\tilde
C}_{\text{e}}(\Delta_{a},x)\;\overline{\mathcal{C}}_{\text{o}}(\Delta_{a},\bar
x)V_{Q/2+iP}(0)\right) \label{LambdaV}%
\end{align}
where $\mathcal{\tilde C}_{\text{e}}(\Delta_{a},x)$ and $\mathcal{\tilde
C}_{\text{o}}(\Delta_{a},x)$ are new chain operators\footnote{These operators
depend also on the dimensions $\Delta_{1}=\Delta_{a_{1}}$ and $\Delta
_{2}=\Delta_{a_{2}}$. For not to overload the notations, this dependence is
not indicated explicitly.}. It is convenient to unify the even and odd ones as
the joint series in integer and half-integer powers of $x$%
\begin{align}
\mathcal{C}(\Delta,x)  & =\mathcal{C}_{\text{e}}(\Delta,x)+\mathcal{C}%
_{\text{o}}(\Delta,x)\label{chains}\\
\mathcal{\tilde C}(\Delta,x)  & =\mathcal{\tilde C}_{\text{e}}(\Delta
,x)+\mathcal{\tilde C}_{\text{o}}(\Delta,x)\nonumber
\end{align}
From (\ref{GV}) one finds
\begin{align}
\mathcal{\tilde C}(\Delta,x)V_{\Delta} & =x^{k}G_{k}\mathcal{C}(\Delta
,x)V_{\Delta}\label{Cheq}\\
\mathcal{C}(\Delta,x)V_{\Delta} & =x^{k}\left(  \Delta+2\Delta_{1}k-\Delta
_{2}-k+x\frac\partial{\partial x}\right)  G_{k}\mathcal{\tilde C}%
(\Delta,x)V_{\Delta}\nonumber
\end{align}
For the coefficients in the level expansion
\begin{equation}
\mathcal{C}(\Delta,x)=\sum_{N}x^{N}\mathcal{C}_{N}(\Delta
)\;;\;\;\;\mathcal{\tilde C}(\Delta,x)=\sum_{N}x^{N}\mathcal{\tilde C}%
_{N}(\Delta)\label{Cn}%
\end{equation}
where $N$ runs over non-negative integer and half-integer numbers (levels),
these relations read ($k\leq N$)
\begin{align}
G_{k}\mathcal{C}_{N}(\Delta)V_{\Delta} & =\mathcal{\tilde C}_{N-k}%
(\Delta)V_{\Delta}\label{Cheqn}\\
G_{k}\mathcal{\tilde C}_{N}(\Delta)V_{\Delta} & =\left(  \Delta+2\Delta
_{1}k-\Delta_{2}+N-k\right)  \mathcal{C}_{N-k}(\Delta)V_{\Delta}\nonumber
\end{align}
This system turns out to be enough to reconstruct the chain operators up to
two overall constants. The latter can be fixed by the conditions
$\mathcal{C}_{0}=1$ and $\mathcal{\tilde C}_{0}=1$. Explicitly one finds,
level by level
\begin{align}
\mathcal{C}_{1/2}(\Delta)  & =\frac1{2\Delta}G_{-1/2}\nonumber\\
\mathcal{C}_{1}(\Delta)  & =\frac{\Delta+\Delta_{1}-\Delta_{2}}{2\Delta}%
L_{-1}\nonumber\\
\mathcal{C}_{3/2}(\Delta)  & =\frac{\Delta+\Delta_{1}-\Delta_{2}+1/2}%
{2\Delta(2\Delta+1)}G_{-1/2}^{3}+\frac{2\left(  \Delta_{1}-\Delta_{2}\right)
}{4\Delta^{2}+2\widehat{c}\Delta-6\Delta+\widehat{c}}O_{-3/2}\nonumber\\
\mathcal{C}_{2}(\Delta)  & =\frac{(\Delta+\Delta_{1}-\Delta_{2})(\Delta
+\Delta_{1}-\Delta_{2}+1)}{4\Delta(2\Delta+1)}L_{-1}^{2}\label{Ccoff}\\
& \ \ +\frac{2\left(  \Delta_{1}-\Delta_{2}\right)  ^{2}+\Delta-(2\Delta
+1)\left(  \Delta_{1}+\Delta_{2}\right)  }{(2\Delta+3)(4\Delta^{2}%
+2\widehat{c}\Delta-6\Delta+\widehat{c})}G_{-1/2}O_{-3/2}\nonumber\\
& \ \ +\frac{3\left(  \Delta_{1}-\Delta_{2}\right)  ^{2}-2\Delta\left(
\Delta_{1}+\Delta_{2}\right)  -\Delta^{2}}{2\Delta(2\Delta+3)(16\Delta
+3\widehat{c}-3)}O_{-2}\nonumber
\end{align}
and
\begin{align}
\mathcal{\tilde C}_{1/2}(\Delta)  & =\frac{\Delta+\Delta_{1}-\Delta_{2}%
}{2\Delta}G_{-1/2}\nonumber\\
\mathcal{\tilde C}_{1}(\Delta)  & =\frac{\Delta+\Delta_{1}-\Delta_{2}%
+1/2}{2\Delta}L_{-1}\nonumber\\
\mathcal{\tilde C}_{3/2}(\Delta)  & =\frac{(\Delta+\Delta_{1}-\Delta
_{2})(\Delta+\Delta_{1}-\Delta_{2}+1)}{2\Delta(2\Delta+1)}G_{-1/2}%
^{3}\nonumber\\
& \ \ +\frac{2\left(  \Delta_{1}-\Delta_{2}\right)  ^{2}+\Delta-(2\Delta
+1)\left(  \Delta_{1}+\Delta_{2}\right)  }{4\Delta^{2}+2\widehat{c}%
\Delta-6\Delta+\widehat{c}}O_{-3/2}\label{Ctcoff}\\
\mathcal{\tilde C}_{2}(\Delta)  & =\frac{(\Delta+\Delta_{1}-\Delta
_{2}+1/2)(\Delta+\Delta_{1}-\Delta_{2}+3/2)}{4\Delta(2\Delta+1)}L_{-1}%
^{2}\nonumber\\
& \ \ +\frac{2\left(  \Delta_{1}-\Delta_{2}\right)  (\Delta+\Delta_{1}%
-\Delta_{2}+3/2)}{(2\Delta+3)(4\Delta^{2}+2\widehat{c}\Delta-6\Delta
+\widehat{c})}G_{-1/2}O_{-3/2}\nonumber\\
& \ \ +\frac{12\left(  \Delta_{1}-\Delta_{2}\right)  ^{2}-\left(
2\Delta+3\right)  \left(  2\Delta-1+4\left(  \Delta_{1}+\Delta_{2}\right)
\right)  }{8\Delta(2\Delta+3)(16\Delta+3\widehat{c}-3)}O_{-2}\nonumber
\end{align}
where we have used the abbreviations
\begin{align}
O_{-3/2}  & =\frac2{2\Delta+1}L_{-1}G_{-1/2}-G_{-3/2}\label{SCF}\\
O_{-2}  & =3L_{-1}^{2}-4\Delta L_{-2}-3G_{-3/2}G_{-1/2}\nonumber
\end{align}

Further coefficients are systematically evaluated level by level as a solution
of (\ref{Cheq}). At each level (\ref{Cheqn}) is a finite dimensional linear
problem, the coefficients being polynomials in $\widehat{c}$, $\Delta$ and the
``external dimensions'' $\Delta_{1}$ and $\Delta_{2}$. The determinant of the
system is the Kac determinant\cite{Kac} of the corresponding level in the
representation of the superconformal algebra. Therefore, the singularities of
$\mathcal{C}(\Delta)$ are (simple in general) poles located at the singular
dimensions $\Delta=\Delta_{m,n}$, where $(m,n)$ is a pair of positive
integers,
\begin{equation}
\Delta_{m,n}=\frac{Q^{2}}8-\frac{\lambda_{m,n}^{2}}2\label{Dmn}%
\end{equation}
and we introduced a notation
\begin{equation}
\lambda_{m,n}=\frac{mb^{-1}+nb}2\label{lmn}%
\end{equation}
In the NS sector, which we only consider here, $m$ and $n$ are either both
even or both odd. At this values in the $SVir$ module over $V_{\Delta}$
appears a singular vector $D_{m,n}V_{m,n}$, a primary field of dimension
$\Delta_{m,n}+mn/2=\Delta_{m,-n}$. Here, as in ref.\cite{shigher}, we denoted
$V_{m,n}=V_{\Delta_{m,n}}$ and introduced a set of ``singular vector
creation'' operators $D_{m,n}$, which are graded polynomials in the generators
$G_{-k}$ and $L_{-n}$ (with the coefficients depending in $\widehat{c}$ (or
$b$) only) such that $G_{k}D_{m,n}V_{m,n}=0$ for every half integer $k>0$.
These operators are supposed to be normalized as in \cite{shigher}
\begin{equation}
D_{m,n}=G_{-1/2}^{mn}+\ldots\label{Dnorm}%
\end{equation}
Below we will need also the ``conjugate'' operator $D_{m,n}^{\dagger}$
defined, as in ref.\cite{shigher} through the following conjugation rules
$L_{n}^{\dagger}=L_{-n}$ and $G_{k}^{\dagger}=G_{-k}$. Apparently
\begin{equation}
D_{m,n}^{\dagger}D_{m,n}V_{\Delta}(0)=r_{m,n}^{\prime}(\Delta-\Delta
_{m,n})V_{\Delta}(0)+O\left(  (\Delta-\Delta_{m,n})^{2}\right)  V_{\Delta
}(0)\label{lognorm}%
\end{equation}
The coefficient $r_{m,n}^{\prime}$ (the ``logarithmic norm'' of the singular
vector) has been explicitly evaluated in \cite{shigher}
\begin{equation}
r_{m,n}^{\prime}=2^{mn-1}\prod_{\substack{k=1-m,\,l=1-n \\(k,l)\neq(0,0),(m,n)
}}^{m,\,n,\,k+l\in2\mathbb{Z}}\lambda_{k,l}\label{rp}%
\end{equation}

The singular vector $D_{m,n}V_{m,n}$, once appeared in the chain vectors
(\ref{chains}), gives rise to its own chains $\mathcal{C}(\Delta
_{m,-n},x)D_{m,n}V_{m,n}$ and $\mathcal{\tilde C}(\Delta_{m,-n},x)D_{m,n}%
V_{m,n}$ which by themselves satisfy the chain equations (\ref{Cheqn}) with
$\Delta=\Delta_{m,n}$. Apparently
\begin{equation}
\operatorname*{res}_{\Delta=\Delta_{m,n}}\mathcal{C}(\Delta,x)=x^{mn/2}%
X_{m,n}\mathcal{C}(\Delta_{m,-n},x)D_{m,n}\label{res}%
\end{equation}
where $X_{m,n}$ are certain coefficients. Since $\mathcal{C}_{\text{e}}%
(\Delta,x)$ with $\mathcal{\tilde C}_{\text{o}}(\Delta,x)$ and $\mathcal{C}%
_{\text{o}}(\Delta,x)$ with $\mathcal{\tilde C}_{\text{e}}(\Delta,x)$ form two
independent systems, we need to treat separately the integer and half-integer
chains. Denote
\begin{align}
\operatorname*{res}_{\Delta=\Delta_{m,n}}\mathcal{C}_{\text{e}}(\Delta,x)  &
=x^{mn/2}\left\{
\begin{array}
[c]{l}%
X_{m,n}^{\text{(e)}}\mathcal{C}_{\text{e}}(\Delta_{m,-n},x)D_{m,n}%
\;\;\;\;\;\;m,n\;\;\;\text{even}\\
X_{m,n}^{\text{(e)}}\mathcal{C}_{\text{o}}(\Delta_{m,-n},x)D_{m,n}%
\;\;\;\;\;\;m,n\;\;\;\text{odd}%
\end{array}
\right. \label{resC}\\
\operatorname*{res}_{\Delta=\Delta_{m,n}}\mathcal{C}_{\text{o}}(\Delta,x)  &
=x^{mn/2}\left\{
\begin{array}
[c]{l}%
X_{m,n}^{\text{(o)}}\mathcal{C}_{\text{o}}(\Delta_{m,-n},x)D_{m,n}%
\;\;\;\;\;\;m,n\;\;\;\text{even}\\
X_{m,n}^{\text{(o)}}\mathcal{C}_{\text{e}}(\Delta_{m,-n},x)D_{m,n}%
\;\;\;\;\;\;m,n\;\;\;\text{odd}%
\end{array}
\right. \nonumber
\end{align}
and let $\mathcal{C}(\Delta,x)$ with $\mathcal{\tilde C}(\Delta,x)$ be
normalized as above. The coefficients $X_{m,n}^{\text{(e)}}$ and
$X_{m,n}^{\text{(o)}}$ are then uniquely defined. By construction they are
polynomials in the external dimensions $\Delta_{1}$ and $\Delta_{2}$. To
describe these coefficients it is convenient to define the following ``fusion
polynomials'' \cite{vbelavin, leshek}
\begin{align}
P_{m,n}^{\text{(e)}}(x)  & =\prod_{\substack{k\in\{1-m,2,m-1\} \\l\in
\{1-n,2,n-1\} }}^{m+n-k-l\operatorname*{mod}4=0}(x-\lambda_{k,l})\label{Peo}\\
P_{m,n}^{\text{(o)}}(x)  & =\prod_{\substack{k\in\{1-m,2,m-1\} \\l\in
\{1-n,2,n-1\} }}^{m+n-k-l\operatorname*{mod}4=2}(x-\lambda_{k,l})\nonumber
\end{align}
Here e.g. $\{1-m,2,m-1\}$ means ``from $1-m$ to $m-1$ with step $2$'', i.e.,
$1-m,3-m,\ldots,m-1$. The degree of these polynomials
\begin{equation}
p_{\text{e,o}}(m,n)=\deg P_{m,n}^{\text{(e,o)}}(x)\label{deg}%
\end{equation}
coincides with the number of multipliers in the products (\ref{Peo})
\begin{align}
p_{\text{e,o}}(m,n)=mn/2\;\;\;\;\;\;\;\;\;\;\;\;\;\;  & \text{at
}m,n\,\,\text{even}\nonumber\\%
\begin{array}
[c]{c}%
\,\,p_{\text{e}}(m,n)=mn/2-1/2\\
\,\,p_{\text{o}}(m,n)=mn/2+1/2
\end{array}
\;\;\;  & \text{at }m,n\,\,\text{odd}\label{pmn}%
\end{align}
In particular, the parity of $P_{m,n}^{\text{(e,o)}}(x)$ is that of the
integer $\,p_{\text{e,o}}(m,n)$. In the current context it turns out
convenient to parameterize the external dimensions in terms of new variables
$\lambda_{i}$ as
\begin{equation}
\Delta_{i}=\frac{Q^{2}}8-\frac{\lambda_{i}^{2}}2\label{li}%
\end{equation}

We will need some more notations. Consider a three point function of formal
chiral fields, like $\left\langle V_{m,n}(\infty)D_{m,n}^{\dagger}%
V_{\Delta_{1}}(1)V_{\Delta_{2}}(0)\right\rangle $ and $\left\langle
V_{m,n}(\infty)D_{m,n}^{\dagger}V_{\Delta_{1}}(1)V_{\Delta_{2}}%
(0)\right\rangle $. Application of the Ward identities, which read in this
context simply as
\begin{align}
\lbrack G_{k},V_{\Delta_{1}}(1)]  & =\Lambda_{\Delta_{1}}(1)\label{GValg}\\
\{G_{k},\Lambda_{\Delta_{1}}(1)\}  & =(2k\Delta_{1}-\Delta_{2}+\Delta
_{m,n}+N)V_{\Delta_{1}}(1)\nonumber
\end{align}
($N$ is some integer or half integer, corresponding to the level of the
descendent of $V_{m,n}(\infty)$) is a purely algebraic procedure and leads to
the relations
\begin{align}
\left\langle V_{m,n}(\infty)D_{m,n}^{\dagger}V_{\Delta_{1}}(1)V_{\Delta_{2}%
}(0)\right\rangle  & =Y_{m,n}^{\text{(e)}}(\Delta_{1},\Delta_{2})\left\{
\begin{array}
[c]{l}%
\left\langle V_{m,n}(\infty)V_{\Delta_{1}}(1)V_{\Delta_{2}}(0)\right\rangle
\;\;\;\text{at }m,n\;\text{even}\\
\left\langle V_{m,n}(\infty)\Lambda_{\Delta_{1}}(1)V_{\Delta_{2}%
}(0)\right\rangle \;\;\;\text{at }m,n\;\text{odd}%
\end{array}
\right. \label{Yeo}\\
\left\langle V_{m,n}(\infty)D_{m,n}^{\dagger}\Lambda_{\Delta_{1}}%
(1)V_{\Delta_{2}}(0)\right\rangle  & =Y_{m,n}^{\text{(o)}}(\Delta_{1}%
,\Delta_{2})\left\{
\begin{array}
[c]{l}%
\left\langle V_{m,n}(\infty)\Lambda_{\Delta_{1}}(1)V_{\Delta_{2}%
}(0)\right\rangle \;\;\;\text{at }m,n\;\text{even}\\
\left\langle V_{m,n}(\infty)V_{\Delta_{1}}(1)V_{\Delta_{2}}(0)\right\rangle
\;\;\,\;\text{at }m,n\;\text{odd}%
\end{array}
\right. \nonumber
\end{align}
Apparently the $Y_{m,n}^{\text{(e,o)}}(\Delta_{1},\Delta_{2})$ are polynomials
in $\Delta_{1}$ and $\Delta_{2}$, the leading order term being generated by
the term in $D_{m,n}$ with maximum number of generators $G_{k}$, i.e., the one
quoted in (\ref{Dnorm}). Thus the degree of $Y_{m,n}^{\text{(e,o)}}(\Delta
_{1},\Delta_{2})$ is $p_{\text{e,o}}(m,n)$ with leading term
\begin{equation}
Y_{m,n}^{\text{(e,o)}}(\Delta_{1},\Delta_{2})=(\Delta_{1}-\Delta
_{2})^{p_{\text{e,o}}(m,n)}+\;\text{lower order terms}\label{Yleading}%
\end{equation}
On the other hand, the standard decoupling consideration requires this
polynomials to be proportional to the product $P_{m,n}^{\text{(e,o)}}%
(\lambda_{1}+\lambda_{2})P_{m,n}^{\text{(e,o)}}(\lambda_{2}-\lambda_{1})$. The
last is apparently a polynomial in $\Delta_{1}$ and $\Delta_{2}$ with leading
order term $(2\Delta_{1}-2\Delta_{2})^{p_{\text{e,o}}(m,n)}$ and therefore
\begin{equation}
Y_{m,n}^{\text{(e,o)}}(\Delta_{1},\Delta_{2})=2^{-p_{\text{e,o}}(m,n)}%
P_{m,n}^{\text{(e,o)}}(\lambda_{1}+\lambda_{2})P_{m,n}^{\text{(e,o)}}%
(\lambda_{2}-\lambda_{1})\label{YPP}%
\end{equation}

Consider first the case $m,n$ even and the following products
\begin{align}
& \ \left\langle V_{\Delta}(\infty)D_{m,n}^{\dagger}\mathcal{C}_{\text{e}%
}(\Delta,x)V_{\Delta}(0)\right\rangle \nonumber\\
\  & =\frac{x^{mn/2}X_{m,n}^{\text{(e)}}}{\Delta-\Delta_{m,n}}\left\langle
V_{\Delta}(\infty)D_{m,n}^{\dagger}\mathcal{C}_{\text{e}}(\Delta
_{m,-n},x)D_{m,n}V_{\Delta}(0)\right\rangle +O(\Delta-\Delta_{m,n}%
)\label{DCee}\\
& \ \ \left\langle V_{\Delta}(\infty)D_{m,n}^{\dagger}G_{1/2}\mathcal{C}%
_{\text{o}}(\Delta,x)V_{\Delta}(0)\right\rangle \nonumber\\
\  & =\ \frac{x^{mn/2}X_{m,n}^{\text{(o)}}}{\Delta-\Delta_{m,n}}\left\langle
V_{\Delta}(\infty)D_{m,n}^{\dagger}G_{1/2}\mathcal{C}_{\text{o}}(\Delta
_{m,-n},x)D_{m,n}V_{\Delta}(0)\right\rangle +O(\Delta-\Delta_{m,n})\nonumber
\end{align}
The estimates $O(\Delta-\Delta_{m,n})$ follows from the observation, that all
non-polar terms in $\mathcal{C}_{\text{e}}(\Delta,x)$ or $G_{1/2}%
\mathcal{C}_{\text{o}}(\Delta,x)$ are orthogonal to $V_{\Delta}(\infty
)D_{m,n}^{\dagger}$ at $\Delta=\Delta_{m,n}$. Similarly, at $m,n$ odd we have
\begin{align}
& \ \left\langle V_{\Delta}(\infty)D_{m,n}^{\dagger}\mathcal{C}_{\text{o}%
}(\Delta,x)V_{\Delta}(0)\right\rangle \nonumber\\
\  & =\frac{x^{mn/2}X_{m,n}^{\text{(o)}}}{\Delta-\Delta_{m,n}}\left\langle
V_{\Delta}(\infty)D_{m,n}^{\dagger}\mathcal{C}_{\text{e}}(\Delta
_{m,-n},x)D_{m,n}V_{\Delta}(0)\right\rangle +O(\Delta-\Delta_{m,n}%
)\label{DCoo}\\
& \ \ \left\langle V_{\Delta}(\infty)D_{m,n}^{\dagger}G_{1/2}\mathcal{C}%
_{\text{e}}(\Delta,x)V_{\Delta}(0)\right\rangle \nonumber\\
\  & =\ \frac{x^{mn/2}X_{m,n}^{\text{(e)}}}{\Delta-\Delta_{m,n}}\left\langle
V_{\Delta}(\infty)D_{m,n}^{\dagger}G_{1/2}\mathcal{C}_{\text{o}}(\Delta
_{m,-n},x)D_{m,n}V_{\Delta}(0)\right\rangle +O(\Delta-\Delta_{m,n})\nonumber
\end{align}
Then the right hand sides are evaluated using (\ref{lognorm}) and compared
with the left hand sides coming from (\ref{Yeo}). This results
in\cite{vbelavin, leshek}
\begin{equation}
X_{m,n}^{\text{(e,o)}}=\frac{Y_{m,n}^{\text{(e,o)}}(\Delta_{1},\Delta_{2}%
)}{r_{m,n}^{\prime}}\label{Xeo}%
\end{equation}
Notice, that a simple change of the notations turns the above consideration to
a proof of the residue formula for the complementary chain $\mathcal{\tilde
C}(\Delta,x)$
\begin{align}
\operatorname*{res}_{\Delta=\Delta_{m,n}}\mathcal{\tilde C}_{\text{e}}%
(\Delta,x)  & =x^{mn/2}\left\{
\begin{array}
[c]{l}%
X_{m,n}^{\text{(o)}}\mathcal{\tilde C}_{\text{e}}(\Delta_{m,-n},x)D_{m,n}%
\;\;\;\;\;\;m,n\;\;\;\text{even}\\
X_{m,n}^{\text{(o)}}\mathcal{\tilde C}_{\text{o}}(\Delta_{m,-n},x)D_{m,n}%
\;\;\;\;\;\;m,n\;\;\;\text{odd}%
\end{array}
\right. \label{resCt}\\
\operatorname*{res}_{\Delta=\Delta_{m,n}}\mathcal{\tilde C}_{\text{o}}%
(\Delta,x)  & =x^{mn/2}\left\{
\begin{array}
[c]{l}%
X_{m,n}^{\text{(e)}}\mathcal{\tilde C}_{\text{o}}(\Delta_{m,-n},x)D_{m,n}%
\;\;\;\;\;\;m,n\;\;\;\text{even}\\
X_{m,n}^{\text{(e)}}\mathcal{\tilde C}_{\text{e}}(\Delta_{m,-n},x)D_{m,n}%
\;\;\;\;\;\;m,n\;\;\;\text{odd}%
\end{array}
\right. \nonumber
\end{align}

These simple analytic properties are inherited by the corresponding
superconformal blocks. Define the odd and even blocks as
\begin{equation}
\mathcal{F}_{\text{e,o}}\left(
\begin{array}
[c]{cc}%
\Delta_{1} & \Delta_{3}\\
\Delta_{2} & \Delta_{4}%
\end{array}
\left|  \Delta\right|  x\right)  =x^{\Delta-\Delta_{1}-\Delta_{2}}\left\langle
V_{\Delta_{4}}(\infty)V_{\Delta_{3}}(1)\mathcal{C}_{\text{e,o}}^{\Delta
_{1},\Delta_{2}}\left(  \Delta,x\right)  )V_{\Delta}(0)\right\rangle
\label{Feo}%
\end{equation}
where again $V_{\Delta}$ are formal (chiral) primary fields and we have
restored the implicit dependence of the chain operator on the external
dimensions. Normalization is fixed by
\begin{align}
\left\langle V_{\Delta_{4}}(\infty)V_{\Delta_{3}}(1)V_{\Delta}%
(0)\right\rangle  & =1\label{N1}\\
\left\langle V_{\Delta_{4}}(\infty)V_{\Delta_{3}}(1)G_{-1/2}V_{\Delta
}(0)\right\rangle  & =1\nonumber
\end{align}
The poles of $\mathcal{C}_{\text{e,o}}^{\Delta_{1},\Delta_{2}}(\Delta)$ turn
to the poles of the blocks, the residues being evaluated similarly (to be more
compact we suppress the external dimensions in the arguments of the blocks)
\begin{align}
\operatorname*{res}_{\Delta=\Delta_{m,n}}\mathcal{F}_{\text{e}}\left(
\Delta,x\right)   & =B_{m,n}^{\text{(e)}}\left(
\begin{array}
[c]{cc}%
\Delta_{1} & \Delta_{3}\\
\Delta_{2} & \Delta_{4}%
\end{array}
\right)  \left\{
\begin{array}
[c]{l}%
\mathcal{F}_{\text{e}}\left(  \Delta_{m,-n},x\right)  \;\;\;\;\text{at
}m,n\;\text{even}\\
\mathcal{F}_{\text{e}}\left(  \Delta_{m,-n},x\right)  \;\;\;\;\text{at
}m,n\;\text{odd}%
\end{array}
\right. \label{resB}\\
\operatorname*{res}_{\Delta=\Delta_{m,n}}\mathcal{F}_{\text{o}}\left(
\Delta,x\right)   & =B_{m,n}^{\text{(o)}}\left(
\begin{array}
[c]{cc}%
\Delta_{1} & \Delta_{3}\\
\Delta_{2} & \Delta_{4}%
\end{array}
\right)  \left\{
\begin{array}
[c]{l}%
\mathcal{F}_{\text{o}}\left(  \Delta_{m,-n},x\right)  \;\;\;\;\text{at
}m,n\;\text{even}\\
\mathcal{F}_{\text{e}}\left(  \Delta_{m,-n},x\right)  \;\;\;\;\,\text{at
}m,n\;\text{odd}%
\end{array}
\right. \nonumber
\end{align}
Matrix elements $\left\langle V_{\Delta_{4}}(\infty)V_{\Delta_{3}}%
(1)D_{m,n}V_{m,n}(0)\right\rangle $ are evaluated by the same algebraic
procedure
\begin{align}
\left\langle V_{\Delta_{4}}(\infty)V_{\Delta_{3}}(1)D_{m,n}V_{m,n}%
(0)\right\rangle  & =Y_{m,n}^{\text{(e)}}(\Delta_{3},\Delta_{4})\left\{
\begin{array}
[c]{l}%
\left\langle V_{\Delta_{4}}(\infty)V_{\Delta_{3}}(1)V_{m,n}(0)\right\rangle
\;\,\text{at }m,n\;\text{even}\\
\langle V_{\Delta_{4}}(\infty)V_{\Delta_{3}}(1)\widehat{V}_{m,n}%
(0)\rangle\;\,\;\text{at }m,n\;\text{odd}%
\end{array}
\right. \label{Y34}\\
\left\langle V_{\Delta_{4}}(\infty)V_{\Delta_{3}}(1)\widehat{D}_{m,n}%
V_{m,n}(0)\right\rangle  & =Y_{m,n}^{\text{(o)}}(\Delta_{3},\Delta
_{4})\left\{
\begin{array}
[c]{l}%
\langle V_{\Delta_{4}}(\infty)V_{\Delta_{3}}(1)\widehat{V}_{m,n}%
(0)\rangle\;\,\,\text{at }m,n\;\text{even}\\
\left\langle V_{\Delta_{4}}(\infty)V_{\Delta_{3}}(1)V_{m,n}(0)\right\rangle
\;\,\text{at }m,n\;\text{odd}%
\end{array}
\right. \nonumber
\end{align}
where $\widehat{D}_{m,n}=G_{-1/2}D_{m,n}$ and $\widehat{V}_{m,n}%
(0)=G_{-1/2}V_{m,n}$. Combining all together we find\cite{vbelavin, leshek}
\begin{equation}
B_{m,n}^{\text{(e,o)}}\left(
\begin{array}
[c]{cc}%
\Delta_{1} & \Delta_{3}\\
\Delta_{2} & \Delta_{4}%
\end{array}
\right)  =\frac{Y_{m,n}^{\text{(e,o)}}(\Delta_{1},\Delta_{2})Y_{m,n}%
^{\text{(e,o)}}(\Delta_{3},\Delta_{4})}{r_{m,n}^{\prime}}\label{Beo}%
\end{equation}
Important symmetries
\begin{align}
B_{m,n}^{\text{(e,o)}}\left(
\begin{array}
[c]{cc}%
\Delta_{1} & \Delta_{3}\\
\Delta_{2} & \Delta_{4}%
\end{array}
\right)   & =B_{m,n}^{\text{(e,o)}}\left(
\begin{array}
[c]{cc}%
\Delta_{3} & \Delta_{1}\\
\Delta_{4} & \Delta_{2}%
\end{array}
\right)  =B_{m,n}^{\text{(e,o)}}\left(
\begin{array}
[c]{cc}%
\Delta_{2} & \Delta_{4}\\
\Delta_{1} & \Delta_{3}%
\end{array}
\right) \label{Rpty}\\
B_{m,n}^{\text{(e,o)}}\left(
\begin{array}
[c]{cc}%
\Delta_{1} & \Delta_{3}\\
\Delta_{2} & \Delta_{4}%
\end{array}
\right)   & =(-)^{p_{\text{e,o}}(m,n)}B_{m,n}^{\text{(e,o)}}\left(
\begin{array}
[c]{cc}%
\Delta_{2} & \Delta_{3}\\
\Delta_{1} & \Delta_{4}%
\end{array}
\right) \nonumber
\end{align}
follow directly from the symmetry properties of the fusion polynomials.

\section{$\widehat{c}$-recursion}

Analytic properties observed in the previous section give rise to convenient
relations for the superconformal blocks $\mathcal{F}_{\text{e}}\left(
\Delta,x\right)  $ and $\mathcal{F}_{\text{o}}(\Delta,x)$ (we suppress
sometimes the explicit dependence on the external dimensions) which allow
their simple recursive evaluation, e.g. as a power series in $x$. The first
way is to consider analytic properties in the central charge $\widehat{c}$ of
the superconformal algebra\cite{vbelavin, leshek}. In this case the Kac
dimensions (\ref{Dmn}) appear as (again in general simple) poles in
$\widehat{c}$ at
\begin{equation}
\widehat{c}=\widehat{c}_{m,n}(\Delta)=5+2T_{m,n}(\Delta)+2T_{m,n}^{-1}%
(\Delta)\label{cmn}%
\end{equation}
where again $(m,n)$ is a pair of natural numbers, both even or both odd, while
$T_{m,n}(\Delta)$ is a root of the quadric (\ref{Dmn}) in $b^{2}$%
\begin{align}
T_{m,n}(\Delta)  & =\frac{1-4\Delta-mn+\sqrt{16\Delta^{2}+8(mn-1)\Delta
+(m-n)^{2}}}{n^{2}-1}\label{Tmn}\\
T_{m,n}^{-1}(\Delta)  & =\frac{1-4\Delta-mn-\sqrt{16\Delta^{2}+8(mn-1)\Delta
+(m-n)^{2}}}{m^{2}-1}\nonumber
\end{align}
For this particular root all singularities corresponding to $m=1$ are sent to
infinity and only the pairs with $m>1$ count. Notice that the root chosen is
non-singular at $n=1$ so that (\ref{Tmn}) is understood as
\begin{equation}
T_{m,1}(\Delta)=\frac{m^{2}-1}{2(1-4\Delta-m)}\label{Tm1}%
\end{equation}
Corresponding residues are read off from those in (\ref{Beo}) when being
expressed in terms of $\Delta$. In the present context it is convenient to use
the symmetry of the polynomials (\ref{Peo}) to make the multipliers in the
residues (\ref{YPP}) explicit polynomials in $\Delta_{1}$ and $\Delta_{2}$. At
$m,n$ even the multipliers in $P_{m,n}^{\text{(e,o)}}(x)$ always enter in
pairs $(x-\lambda_{k,l})(x-\lambda_{-k,-l})$ and one can ``fold'' the product
$2^{-p_{\text{e,o}}(m,n)}P_{m,n}^{\text{(e,o)}}(\lambda_{1}-\lambda
_{2})P_{m,n}^{\text{(e,o)}}(\lambda_{1}+\lambda_{2})$ as follows
\begin{equation}
Y_{m,n}^{\text{(e,o)}}(\Delta_{1},\Delta_{2})=\prod_{\substack{k\in\{1,2,m-1\}
\\l\in\{1-n,2,n-1\} }}^{m+n-k-l\operatorname*{mod}4=0,2}\mathcal{Y}%
_{k,l}^{(m,n)}(\Delta_{1},\Delta_{2},\Delta)\label{PP}%
\end{equation}
where
\begin{equation}
\mathcal{Y}_{k,l}^{(m,n)}(\Delta_{1},\Delta_{2},\Delta)=(\Delta_{1}-\Delta
_{2})^{2}+\Lambda_{k,l}^{(m,n)}(\Delta)(\Delta_{1}+\Delta_{2}-\Lambda
_{1,1}^{(m,n)}(\Delta))+\frac14\left(  \Lambda_{k,l}^{(m,n)}(\Delta)\right)
^{2}\label{Ykl}%
\end{equation}
and
\begin{equation}
\Lambda_{k,l}^{(m,n)}(\Delta)=\frac{k^{2}T_{m,n}^{-1}(\Delta)+2kl+l^{2}%
T_{m,n}(\Delta)}4\label{Lkl}%
\end{equation}
Similar folding is possible also at $m,n$ odd if the degree of $P_{m,n}$ is
even. If it is odd, the term with $(k,l)=(0,0)$ leads to an extra multiplier
$\Delta_{1}-\Delta_{2}$.

Once the multipliers in (\ref{Beo}) are expressed in terms of $\Delta$, the
corresponding residues in $\widehat{c}$ read
\begin{align}
\operatorname*{res}_{\widehat{c}=\widehat{c}_{m,n}}\mathcal{F}_{\text{e}%
}\left(  \widehat{c},\Delta,x\right)   & =\widehat{B}_{m,n}^{\text{(e)}%
}(\Delta)\left\{
\begin{array}
[c]{l}%
\mathcal{F}_{\text{e}}\left(  \widehat{c}_{m,n},\Delta+mn/2,x\right)
\;\;\;\;\text{at }m,n\;\text{even}\\
\mathcal{F}_{\text{o}}\left(  \widehat{c}_{m,n},\Delta+mn/2,x\right)
\;\;\;\;\text{at }m,n\;\text{odd}%
\end{array}
\right. \label{rescF}\\
\operatorname*{res}_{\widehat{c}=\widehat{c}_{m,n}}\mathcal{F}_{\text{o}%
}\left(  \widehat{c},\Delta,x\right)   & =\widehat{B}_{m,n}^{\text{(o)}%
}(\Delta)\left\{
\begin{array}
[c]{l}%
\mathcal{F}_{\text{o}}\left(  \widehat{c}_{m,n},\Delta+mn/2,x\right)
\;\;\;\;\text{at }m,n\;\text{even}\\
\mathcal{F}_{\text{e}}\left(  \widehat{c}_{m,n},\Delta+mn/2,x\right)
\;\;\;\;\,\text{at }m,n\;\text{odd}%
\end{array}
\right. \nonumber
\end{align}
where
\begin{equation}
\widehat{B}_{m,n}^{\text{(e,o)}}(\Delta)=B_{m,n}^{\text{(e.o)}}(\Delta
)\frac{16(T_{m,n}(\Delta)-T_{m,n}^{-1}(\Delta))}{(n^{2}-1)T_{m,n}%
(\Delta)-(m^{2}-1)T_{m,n}^{-1}(\Delta)}\label{Bmn}%
\end{equation}
the last fraction corresponding to $-\partial\widehat{c}/\partial\Delta$ along
the $(m,n)$ Kac quadric (\ref{Dmn}).

The asymptotic of $\mathcal{F}_{\text{e,o}}\left(  \widehat{c},\Delta
,x\right)  $ at $\widehat{c}\rightarrow\infty$ has been recovered in
\cite{vbelavin, leshek}. Analytic properties in $\widehat{c}$ sum up to the
following relations%

\begin{align}
\mathcal{F}_{\text{e,o}}\left(  \widehat{c},\Delta,x\right)  =f_{\text{e,o}%
}(\Delta,x)  & +\sum_{m,n\;\text{even}}\frac{\widehat{B}_{m,n}^{\text{(e,o)}%
}(\Delta)}{\widehat{c}-\widehat{c}_{m,n}(\Delta)}\mathcal{F}_{\text{e,o}%
}(\widehat{c}_{m,n},\Delta+mn/2,x)\label{crec}\\
& \ \ +\sum_{\substack{m,n\;\text{odd} \\m>1 }}\frac{\widehat{B}%
_{m,n}^{\text{(e,o)}}(\Delta)}{\widehat{c}-\widehat{c}_{m,n}(\Delta
)}\mathcal{F}_{\text{o,e}}(\widehat{c}_{m,n},\Delta+mn/2,x)\nonumber
\end{align}
where%

\begin{align}
f_{\text{e}}(\Delta,x)  & =x^{\Delta-\Delta_{1}-\Delta_{2}}{}_{2}F_{1}%
(\Delta+\Delta_{1}-\Delta_{2},\Delta+\Delta_{3}-\Delta_{4},2\Delta
,x)\label{feo}\\
f_{\text{o}}(\Delta,x)  & =\frac1{2\Delta}x^{\Delta-\Delta_{1}-\Delta_{2}%
+1/2}{}_{2}F_{1}\left(  \Delta+\Delta_{1}-\Delta_{2}+\frac12,\Delta+\Delta
_{3}-\Delta_{4}+\frac12,2\Delta+1,x\right) \nonumber
\end{align}

Equations (\ref{crec}) apparently can be used for recursive evaluation of the
coefficients in the series expansions
\begin{equation}
\mathcal{F}_{\text{e,o}}\left(  \widehat{c},\Delta,x\right)  =x^{\Delta
-\Delta_{1}-\Delta_{2}}\sum_{k}x^{N}F_{\text{e,o}}^{(N)}\left(  \widehat
{c},\Delta\right) \label{FN}%
\end{equation}
where the sum is over non-negative integer (for $\mathcal{F}_{\text{e}}$) or
half-integer (for $\mathcal{F}_{\text{o}}$) numbers
\begin{align}
F_{\text{e,o}}^{(N)}\left(  \widehat{c},\Delta\right)  =f_{\text{e,o}}%
^{(N)}(\Delta)  & +\sum_{m,n\;\text{even}}^{mn/2\leq N}\frac{\widehat{B}%
_{m,n}^{\text{(e,o)}}(\Delta)}{\widehat{c}-\widehat{c}_{m,n}(\Delta
)}F_{\text{e,o}}^{(N-mn/2)}\left(  \widehat{c}_{m,n},\Delta+mn/2\right)
\label{FNrec}\\
& +\sum_{\substack{m,n\;\text{odd} \\m>1 }}^{mn/2\leq N}\frac{\widehat
{B}_{m,n}^{\text{(e,o)}}(\Delta)}{\widehat{c}-\widehat{c}_{m,n}(\Delta
)}F_{\text{o,e}}^{(N-mn/2)}\left(  \widehat{c}_{m,n},\Delta+mn/2\right)
\nonumber
\end{align}
In (\ref{FNrec})
\begin{align}
f_{\text{e}}^{(N)}(\Delta)  & =\frac{(\Delta+\Delta_{1}-\Delta_{2})_{N}%
(\Delta+\Delta_{3}-\Delta_{4})_{N}}{N!(2\Delta)_{N}}\label{fN}\\
f_{\text{o}}^{(N)}(\Delta)  & =\frac{(\Delta+\Delta_{1}-\Delta_{2}%
+1/2)_{N-1/2}(\Delta+\Delta_{3}-\Delta_{4}+1/2)_{N-1/2}}{(N-1/2)!(2\Delta
)_{N+1/2}}\nonumber
\end{align}

\section{Elliptic recursion}

For practical calculations another relation turns out to be much more
convenient. This relation is called the elliptic recursion (since it requires
a parametrization in terms of the elliptic functions) or, sometimes, the
$\Delta$-recursion, because it is based on the analytic properties of the
blocks in $\Delta$ instead of $\widehat{c}$. As in ref.\cite{dblock} we
understand the variable $x$ as the modulus of the elliptic curve
$y^{2}=t(1-t)(1-xt)$ and introduce the ratio of its periods $\tau$
\begin{equation}
\tau=i\frac{K(1-x)}{K(x)}\label{tau}%
\end{equation}
where
\begin{equation}
K(x)=\int_{0}^{1}\frac{dt}{2y(t)}\label{K}%
\end{equation}
is the complete elliptic integral of the first kind. Let also $q=\exp(i\pi
\tau)$ and denote in the standard way
\begin{equation}
\theta_{3}(q)=\sum_{n=-\infty}^{\infty}q^{n^{2}}\,;\;\;\;\;\;\theta
_{2}(q)=\sum_{n=-\infty}^{\infty}q^{(n+1/2)^{2}}\label{theta}%
\end{equation}
so that
\begin{equation}
x=\frac{\theta_{2}^{4}(q)}{\theta_{3}^{4}(q)}\label{xq}%
\end{equation}
inverts (\ref{tau}). This elliptic parametrization has important advantages.
Eq.(\ref{xq}) maps the half plane $\operatorname*{Im}\tau>0$ to the universal
covering of the $x$-plane with punctures at $0$, $1$ and $\infty$. The power
expansions (\ref{FN}) of the blocks in $x$ converge inside the disk $\left|
x\right|  <1$. Once reexpanded in $q$ it converges inside $\left|  q\right|
<1$, i.e., on the whole covering and therefore gives there a uniform
approximation. Naturally, even in the region $\left|  x\right|  <1$ it is
expected to converge faster. The elliptic recursion, which we describe now,
gives the blocks directly in terms of the elliptic variable and allows to
generate the $q$-series in a simple way.

Define the ``elliptic'' blocks $H_{\text{e,o}}(\Delta,q)$ through the
relations
\begin{align}
\mathcal{F}_{\text{e}}\left(
\begin{array}
[c]{cc}%
\Delta_{1} & \Delta_{3}\\
\Delta_{2} & \Delta_{4}%
\end{array}
\left|  \Delta\right|  x\right)   & =(16q)^{\Delta-Q^{2}/8}\frac
{x^{Q^{2}/8-\Delta_{1}-\Delta_{2}}(1-x)^{Q^{2}/8-\Delta_{1}-\Delta_{3}}%
}{\theta_{3}^{4\sum_{i=1}^{4}\Delta_{i}-3Q^{2}/2}(q)}H_{\text{e}}\left(
\begin{array}
[c]{cc}%
\lambda_{1} & \lambda_{3}\\
\lambda_{2} & \lambda_{4}%
\end{array}
\left|  \Delta\right|  q\right) \label{FH}\\
2\mathcal{F}_{\text{o}}\left(
\begin{array}
[c]{cc}%
\Delta_{1} & \Delta_{3}\\
\Delta_{2} & \Delta_{4}%
\end{array}
\left|  \Delta\right|  x\right)   & =(16q)^{\Delta-Q^{2}/8}\frac
{x^{Q^{2}/8-\Delta_{1}-\Delta_{2}}(1-x)^{Q^{2}/8-\Delta_{1}-\Delta_{3}}%
}{\theta_{3}^{4\sum_{i=1}^{4}\Delta_{i}-3Q^{2}/2}(q)}H_{\text{o}}\left(
\begin{array}
[c]{cc}%
\lambda_{1} & \lambda_{3}\\
\lambda_{2} & \lambda_{4}%
\end{array}
\left|  \Delta\right|  q\right) \nonumber
\end{align}
where the parametrization (\ref{li}) of the external dimensions $\Delta_{i}$
in terms of $\lambda_{i}$ is implied. In order, the elliptic blocks satisfy
the following relations (the elliptic recursion)%

\begin{align}
H_{\text{e}}(\Delta,q)=\theta_{3}(q^{2})  & +\sum_{m,n\;\text{even}}%
\frac{q^{mn/2}R_{m,n}^{\text{(e)}}}{\Delta-\Delta_{m,n}}H_{\text{e}}\left(
\Delta_{m,-n},q\right)  +\sum_{m,n\;\text{odd}}\frac{q^{mn/2}R_{m,n}%
^{\text{(e)}}}{\Delta-\Delta_{m,n}}H_{\text{o}}\left(  \Delta_{m,-n},q\right)
\label{Hrec}\\
H_{\text{o}}(\Delta,q)  & =\sum_{m,n\;\text{even}}\frac{q^{mn/2}%
R_{m,n}^{\text{(o)}}}{\Delta-\Delta_{m,n}}H_{\text{o}}\left(  \Delta
_{m,-n},q\right)  +\sum_{m,n\;\text{odd}}\frac{q^{mn/2}R_{m,n}^{\text{(o)}}%
}{\Delta-\Delta_{m,n}}H_{\text{e}}\left(  \Delta_{m,-n},q\right) \nonumber
\end{align}
the residues reading simply
\begin{equation}
R_{m,n}^{\text{(e,o)}}=\frac{P_{m,n}^{\text{(e,o)}}(\lambda_{1}+\lambda
_{2})P_{m,n}^{\text{(e,o)}}(\lambda_{1}-\lambda_{2})P_{m,n}^{\text{(e,o)}%
}(\lambda_{3}+\lambda_{4})P_{m,n}^{\text{(e,o)}}(\lambda_{3}-\lambda_{4}%
)}{r_{m,n}^{\prime}}\label{Reo}%
\end{equation}
These relations take into account the analytic properties of the
superconformal blocks in $\Delta$ described in section 3. In addition they
imply that in the limit $\Delta\rightarrow\infty$%
\begin{align}
H_{\text{e}}(\Delta,q)  & =\theta_{3}(q^{2})+O(\Delta^{-1})\label{Hass}\\
H_{\text{o}}(\Delta,q)  & =\Delta^{-1}\theta_{2}(q^{2})+O(\Delta
^{-2})\nonumber
\end{align}
The first asymptotic is plugged explicitly into the relations, while the
second is automatically generated by the recursion. To derive the relations
(\ref{Hrec}) we need to justify the $\Delta\rightarrow\infty$ asymptotic of
the blocks, summed up in (\ref{FH}) and (\ref{Hass}). The arguments will be
reported elsewhere.

Like the $\widehat{c}$-recursion of section 4, relations (\ref{Hrec}) allow to
evaluate recursively the series expansions of the elliptic blocks in powers of
$q$%
\begin{equation}
H_{\text{e.o}}(\Delta,q)=\sum_{N}q^{N}h_{\text{e,o}}^{(N)}(\Delta)\label{hN}%
\end{equation}
Again $H_{\text{e}}(\Delta,q)$ expands in non-negative integer powers of $q$
while $H_{\text{o}}(\Delta,q)$ is a series in positive half-integer ones.
Relations (\ref{Hrec}) give%

\begin{equation}
h_{\text{e,o}}^{(N)}(\Delta)=\eta_{\text{e,o}}^{(N)}+\sum_{m,n\;\text{even}%
}^{mn/2\leq N}\frac{R_{m,n}^{\text{(e,o)}}h_{\text{e,o}}^{(N-mn/2)}%
(\Delta_{m,-n})}{\Delta-\Delta_{m,n}}+\sum_{m,n\;\text{odd}}^{mn/2\leq N}%
\frac{R_{m,n}^{\text{(e,o)}}h_{\text{o,e}}^{(N-mn/2)}\Delta_{m,-n})}%
{\Delta-\Delta_{m,n}}\label{HNrc}%
\end{equation}
where $\eta_{\text{o}}^{(N)}=0$ and $\eta_{\text{e}}^{(N)}$ are coefficients
in the $q$-expansion of $\theta_{3}(q^{2})$. In practice this relation allows
a much better algorithm as compared to the $\widehat{c}$-recursion
(\ref{FNrec}), mostly because $\widehat{c}$ is fixed and the values
$R_{m,n}^{\text{(e,o)}}$, $\Delta_{m,n}$ remain the same at all iteration
steps (unlike (\ref{FNrec}), where at each level they have to be recomputed
for the ``shifted'' values of $\Delta$). Once a necessary number of residues
and dimensions is available, the remaining recursive procedure runs very fast.

\section{Superconformal bootstrap}

With the structure constants (\ref{C3}) and superconformal blocks known one is
in the position to evaluate the four-point function (\ref{schan}) of basic NS
fields in SLFT. The goal of this section is to verify numerically the crossing
symmetry relations (\ref{cross}).

For the reasons discussed above, we will use the elliptic representation
(\ref{FH}) of the blocks. The four-point function (\ref{schan}) acquires the
form
\begin{equation}
G\left(  \left.
\begin{array}
[c]{cc}%
a_{1} & a_{3}\\
a_{2} & a_{4}%
\end{array}
\right|  x,\bar x\right)  =\frac{(x\bar x)^{Q^{2}/8-\Delta_{1}-\Delta_{2}%
}[(1-x)(1-\bar x)]^{Q^{2}/8-\Delta_{1}-\Delta_{3}}}{[\theta_{3}(q)\theta
_{3}(\bar q)]^{4\sum_{i}\Delta_{i}-3Q^{2}/2}}g\left(  \left.
\begin{array}
[c]{cc}%
a_{1} & a_{3}\\
a_{2} & a_{4}%
\end{array}
\right|  \tau,\bar\tau\right) \label{G}%
\end{equation}
where
\begin{align}
\  & g\left(  \left.
\begin{array}
[c]{cc}%
a_{1} & a_{3}\\
a_{2} & a_{4}%
\end{array}
\right|  \tau,\bar\tau\right)  =\label{gtau}\\
\ \ \  & \int\frac{dP}{4\pi}\left|  16q\right|  ^{P^{2}}\left[  \mathbb{C}%
_{a_{1},a_{2}}^{Q/2+iP}\mathbb{C}_{a_{3},a_{4}}^{Q/2-iP}H_{\text{e}}\left(
\Delta,q\right)  H_{\text{e}}\left(  \Delta,\bar q\right)  -\mathbb{\tilde
C}_{a_{1},a_{2}}^{Q/2+iP}\mathbb{\tilde C}_{a_{3},a_{4}}^{Q/2-iP}H_{\text{o}%
}\left(  \Delta,q\right)  H_{\text{o}}\left(  \Delta,\bar q\right)  \right]
\nonumber
\end{align}
and $\Delta=Q^{2}/8+P^{2}/2$. The first of the relations (\ref{cross}) is
verified analytically. Indeed, the identities
\begin{align}
H_{\text{e}}\left(
\begin{array}
[c]{cc}%
\lambda_{1} & \lambda_{3}\\
\lambda_{2} & \lambda_{4}%
\end{array}
\left|  \Delta\right|  -q\right)   & =H_{\text{e}}\left(
\begin{array}
[c]{cc}%
\lambda_{1} & \lambda_{4}\\
\lambda_{2} & \lambda_{3}%
\end{array}
\left|  \Delta\right|  q\right) \label{Hs}\\
H_{\text{o}}\left(
\begin{array}
[c]{cc}%
\lambda_{1} & \lambda_{3}\\
\lambda_{2} & \lambda_{4}%
\end{array}
\left|  \Delta\right|  e^{i\pi}q\right)   & =e^{i\pi/2}H_{\text{o}}\left(
\begin{array}
[c]{cc}%
\lambda_{1} & \lambda_{3}\\
\lambda_{2} & \lambda_{4}%
\end{array}
\left|  \Delta\right|  q\right) \nonumber
\end{align}
for the elliptic blocks (the latter are easily derived from (\ref{Hrec}) and
the symmetries (\ref{Rpty}) of the residues) directly result in
\begin{equation}
g\left(  \left.
\begin{array}
[c]{cc}%
a_{1} & a_{3}\\
a_{2} & a_{4}%
\end{array}
\right|  \tau,\bar\tau\right)  =g\left(  \left.
\begin{array}
[c]{cc}%
a_{1} & a_{4}\\
a_{2} & a_{3}%
\end{array}
\right|  \tau+1,\bar\tau+1\right) \label{c1}%
\end{equation}

The second relation in eq.(\ref{cross}) in terms of the function $g$ reads
\begin{equation}
g\left(  \left.
\begin{array}
[c]{cc}%
a_{1} & a_{3}\\
a_{2} & a_{4}%
\end{array}
\right|  \tau,\bar\tau\right)  =(\tau\bar\tau)^{3Q^{2}/4-2\sum_{i}\Delta_{i}%
}g\left(  \left.
\begin{array}
[c]{cc}%
a_{1} & a_{2}\\
a_{3} & a_{4}%
\end{array}
\right|  -\frac1\tau,-\frac1{\bar\tau}\right) \label{c2}%
\end{equation}
It is a difficult mathematical problem to recover this property from the
representation (\ref{g}). However, with the fast algorithms based on the
elliptic representation of the blocks, it is an affordable problem for
numerical analysis.

As a simplest numerical test we have chosen the external parameters
$a_{1}=a_{2}=a_{3}=a_{4}=Q/2$. The structure constants (\ref{C3}) vanish for
these values of the external parameters, so that we take a first derivative in
all four of them. This means that we consider the four point function of the
primary fields $V_{Q/2}^{\prime}=\phi\exp(Q\phi/2)$. Denoting
\begin{equation}
\frac{\partial^{4}}{\partial a_{1}\partial a_{2}\partial a_{3}\partial a_{4}%
}\left.  g\left(  \left.
\begin{array}
[c]{cc}%
a_{1} & a_{3}\\
a_{2} & a_{4}%
\end{array}
\right|  \tau,\bar\tau\right)  \right|  _{a_{i}=Q/2}=4\Upsilon_{\text{NS}%
}^{\prime4}(0)\Upsilon_{\text{R}}^{2}(b)f(\tau,\bar\tau)\label{gf}%
\end{equation}
we find
\begin{equation}
f(\tau,\bar\tau)=\int\frac{dP}{4\pi}\left|  16q\right|  ^{P^{2}}\left[
r_{\text{e}}(P)\left|  H_{\text{e}}\left(
\begin{array}
[c]{cc}%
0 & 0\\
0 & 0
\end{array}
\left|  \Delta\right|  q\right)  \right|  ^{2}+r_{\text{o}}(P)\left|
H_{\text{o}}\left(
\begin{array}
[c]{cc}%
0 & 0\\
0 & 0
\end{array}
\left|  \Delta\right|  q\right)  \right|  ^{2}\right] \label{f}%
\end{equation}
The auxiliary functions $r_{\text{e}}(P)$ and $r_{\text{o}}(P)$ read in terms
of standard upsilon functions $\Upsilon_{b}(x)$
\begin{align}
r_{\text{e}}(P)  & =\frac{\Upsilon_{b}(iP)\Upsilon_{b}(-iP)\Upsilon_{b}%
^{2}(iP+Q/2)}{\Upsilon_{b}^{8}(Q/4+iP/2)\Upsilon_{b}^{8}(Q/4-iP/2)}%
\label{reo}\\
r_{\text{o}}(P)  & =\frac{\Upsilon_{b}(iP)\Upsilon_{b}(-iP)\Upsilon_{b}%
^{2}(Q/2+iP)}{\Upsilon_{b}^{8}(Q/4+b/2+iP/2)\Upsilon^{8}(Q/4+b^{-1}%
/2+iP/2)}\nonumber
\end{align}
and allow the following integral representations
\begin{align}
r_{\text{e}}(P)  & =P^{2}\exp\left\{  \int_{0}^{\infty}\frac{dt}t\left[
-\frac{(1+6b^{2}+b^{4})e^{-t}}{2b^{2}}\right.  \right. \label{re}\\
& \ \ \ \ \ +\left.  \left.  \frac{8\cos(Pt/2)\cosh[(b+1/b)t/4]-2\cos
(Pt)\cosh^{2}[(b-1/b)t/4]-6}{\sinh(t/2b)\sinh(bt/2)}\right]  \right\}
\nonumber
\end{align}%
\begin{align}
r_{\text{o}}(P)  & =P^{2}\exp\left\{  \int_{0}^{\infty}\frac{dt}t\left[
-\frac{(1-2b^{2}+b^{4})e^{-t}}{2b^{2}}\right.  \right. \label{ro}\\
& \ \ \ \ \ \left.  \left.  +\frac{8\cos(Pt/2)\cosh[(b-1/b)t/4]-2\cos
(Pt)\cosh^{2}[(b-1/b)t/4]-6}{\sinh(t/2b)\sinh(bt/2)}\right]  \right\}
\nonumber
\end{align}%

\begin{figure}
[tbh]
\begin{center}
\includegraphics[
height=3.5198in,
width=5.0194in
]%
{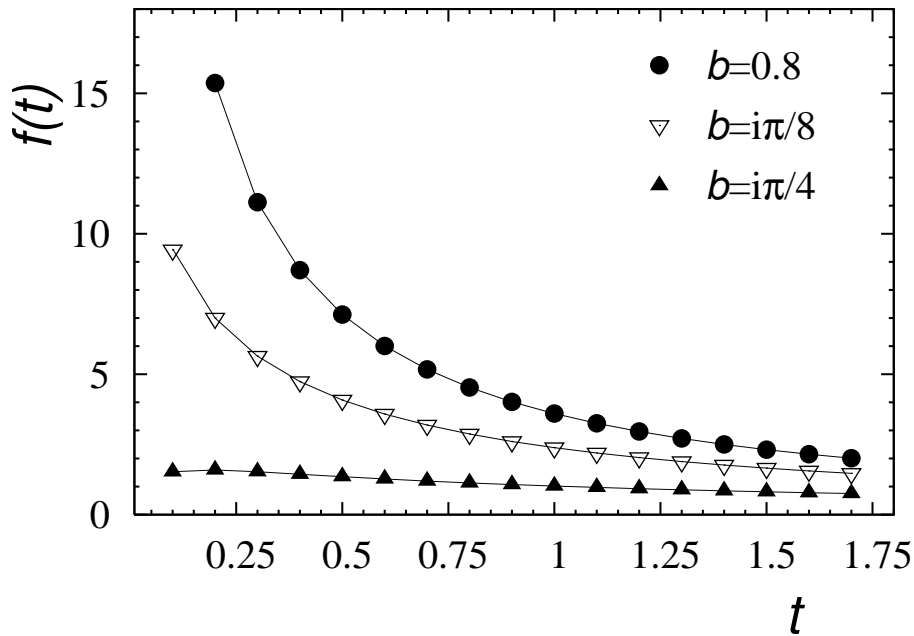}%
\caption{The results of comparison for different values of $b$.}%
\end{center}
\end{figure}

In Figure 1 the values of $f(\tau,\bar\tau)$ (solid curves) are compared with
that of $(\tau\bar\tau)^{-Q^{2}/4}\times f(-1/\tau,-1/\bar\tau)$ (symbols) for
pure imaginary $\tau=it$ and certain values of the parameter $b$. To give an
idea about the accuracy, some numbers are shown in the Table 1. These numbers
correspond to the approximation of the elliptic blocks $H_{\text{e}}(q)$ and
$q^{1/2}H_{\text{o}}(q)$ as power series up to the order $q^{6}$. It seems
like the main source of the discrepancy is in this approximation. For example
at $t=0.5$ ($q^{2}\approx0.0432$) the numbers differ in the fifth decimal
digit (e.g., the numbers $7.12534$ and $7.12512$ in the Table), while the
approximation improved up to the order $q^{10}$ gives seven correct decimal
digits (respectively $7.12511575$ and $7.12511599 $).

\begin{center}%
\begin{table}[tbp] \centering
\begin{tabular}
[c]{|c|c|c|c|c|}\hline
& $f(t)$ & $t^{-Q^{2}/2}f(1/t)$ & $f(t)$ & $t^{-Q^{2}/2}f(1/t)$\\
$t$ & for $b=0.8$ & for $b=0.8$ & for $b=i\pi/4$ & for $b=i\pi/4$\\\hline
0.1 & 25.900821 & 25.456246 & 1.5305811 & 1.5232342\\
0.2 & 15.372272 & 15.365205 & 1.5884637 & 1.5881483\\
0.3 & 11.123984 & 11.123971 & 1.5267458 & 1.5264759\\
0.4 & 8.7059097 & 8.7060123 & 1.4398383 & 1.4396855\\
0.5 & 7.1251188 & 7.125178 & 1.3522805 & 1.3521968\\
0.6 & 6.0067008 & 6.0067341 & 1.2708572 & 1.2708109\\
0.7 & 5.1733043 & 5.1733228 & 1.1971168 & 1.1970912\\
0.8 & 4.528805 & 4.5288147 & 1.1308767 & 1.1308632\\
0.9 & 4.0160968 & 4.0161008 & 1.0713992 & 1.0713936\\\hline
\end{tabular}
\caption{Numerical data for $f(t)$ for $b=0.8$ and $b=i\pi/4$.\label{key}}%
\end{table}%
\end{center}

We have performed a similar comparison for some different values of
$a_{1},a_{2},a_{3},a_{4}$ (chosen at random, but close enough to $Q/2$ to
preserve the convergence of the representations (\ref{re}) and (\ref{ro})).
E.g., at
\begin{equation}
a_{1}=\frac9{14}Q\,;\;a_{3}=\frac{2Q}3\,;\;a_{2}=\frac{13Q}{30}\,;\;a_{4}%
=\frac{7Q}{10}\label{atest}%
\end{equation}
we have for $b=2.8$ and $t=0.3$
\begin{align}
f_{a_{1},a_{2},a_{3},a_{4}}(t)  & =2141.5325\label{ftest}\\
t^{3Q^{2}/2-4\sum_{i}\Delta_{i}}f_{a_{1},a_{3},a_{2},a_{4}}(1/t)  &
=2141.5101\nonumber
\end{align}

\section{Discussion}

In the present paper a preliminary analysis of the bootstrap properties has
been performed for the four point functions in the $N=1$ supersymmetric
Liouville field theory. Some, mostly numerical, arguments are presented that
the NS operator algebra, based on the structure constants (\ref{C3}),
satisfies the locality property. Of course, a separate analysis is needed to
include the Ramond sector. This problem remains for future work.

Even in the purely NS sector the program is not yet completely finished. We
considered only the four point functions of four ``bottom'' components $V$ of
any supermultiplet. Correlation functions involving other components, like
$\left\langle VVVW\right\rangle $ or $\left\langle \Lambda\Lambda
VV\right\rangle $ etc., remain to be studied. Obviously they can be expressed
in terms of the structure constants (\ref{C3}), like in eq.(\ref{schan}) for
$\left\langle VVVV\right\rangle $, with different superconformal blocks. We
arrive at the problem to describe completely the set of $32$ different blocks
like (in obvious notations)
\begin{align}
\mathcal{F}_{\text{e,o}}\left(
\begin{array}
[c]{cc}%
\Delta_{1} & \widehat{\Delta}_{3}\\
\Delta_{2} & \Delta_{4}%
\end{array}
\left|  \Delta\right|  x\right)   & =\left\langle V_{4}(\infty)\Lambda
_{3}(1)\left|  \Delta,\widehat{\Delta}\right|  V_{1}(x)V_{2}(0)\right\rangle
\label{Fex}\\
\mathcal{F}_{\text{e,o}}\left(
\begin{array}
[c]{cc}%
\widehat{\Delta}_{1} & \widehat{\Delta}_{3}\\
\Delta_{2} & \Delta_{4}%
\end{array}
\left|  \Delta\right|  x\right)   & =\left\langle V_{4}(\infty)\Lambda
_{3}(1)\left|  \Delta,\widehat{\Delta}\right|  \Lambda_{1}(x)\Lambda
_{2}(0)\right\rangle \nonumber\\
& \ \ \text{etc.}\nonumber
\end{align}
where in the present context $V$ and $\Lambda$ are formal chiral components of
the ``right'' supermultiplet, and, by definition
\begin{equation}
\left\langle V_{4}(\infty)\Lambda_{3}(1)\left|  \Delta,\widehat{\Delta
}\right|  V_{1}(x)V_{2}(0)\right\rangle =\left\langle V_{4}(\infty)\Lambda
_{3}(1)\mathcal{C}_{\text{e,o}}^{\Delta_{1},\Delta_{2}}(\Delta,x)V_{\Delta
}(0)\right\rangle \label{VL}%
\end{equation}
(similarly for other components).

The problem is slightly less involved than it seems at first glance. A study
of the superprojective Ward identities, similar to that in Appendix A, allows
to reduce the set to $8$ independent functions. Moreover, their analytic
properties follow directly form those of the basic chains $\mathcal{C}%
_{\text{e,o}}(\Delta,x)$ and $\mathcal{\tilde C}_{\text{e,o}}(\Delta,x)$. In
particular, the residues at the singular dimensions $\Delta=\Delta_{m,n}$ are
expressed similarly to (\ref{Beo}) in terms of the fusion polynomials.
Construction of the recursive representations, analogous to (\ref{crec}) or
(\ref{Hrec}), presents a separate problem, which we hope to analyze in the future.

Another topic of a future report is the justifications of the asymptotic
(\ref{FH}) and (\ref{Hass}) of the block at $\Delta\rightarrow\infty$, which
has been simply conjectured in section 5 in order to write down the elliptic
recursion relations. This analysis requires several essential steps and
presently we prefer to skip it.

The superconformal blocks, considered above, are expected to satisfy certain
``crossing'', or ``fusion'' relations, the set with different values of the
intermediate dimension forming an infinite dimensional representation of the
monodromy group. In the present case the group is equivalent to the modular
group generated by the maps $x\rightarrow x/(x-1)$ and $x\rightarrow1-x$. In
terms of the elliptic parameter $\tau$ these are $\tau\rightarrow\tau+1$ and
$\tau\rightarrow-1/\tau$. The representation is completely determined by the
relations
\begin{align}
\mathcal{F}_{\text{e}}\left(
\begin{array}
[c]{cc}%
\Delta_{1} & \Delta_{3}\\
\Delta_{2} & \Delta_{4}%
\end{array}
\left|  \Delta\right|  x\right)   & =e^{-i\pi(\Delta-\Delta_{1}-\Delta_{2}%
)}(1-x)^{-2\Delta_{1}}\mathcal{F}_{\text{e}}\left(
\begin{array}
[c]{cc}%
\Delta_{1} & \Delta_{3}\\
\Delta_{2} & \Delta_{4}%
\end{array}
\left|  \Delta\right|  \frac{xe^{i\pi}}{1-x}\right) \label{crossI}\\
\mathcal{F}_{\text{o}}\left(
\begin{array}
[c]{cc}%
\Delta_{1} & \Delta_{3}\\
\Delta_{2} & \Delta_{4}%
\end{array}
\left|  \Delta\right|  x\right)   & =e^{-i\pi(\Delta-\Delta_{1}-\Delta
_{2}+1/2)}(1-x)^{-2\Delta_{1}}\mathcal{F}_{\text{o}}\left(
\begin{array}
[c]{cc}%
\Delta_{1} & \Delta_{3}\\
\Delta_{2} & \Delta_{4}%
\end{array}
\left|  \Delta\right|  \frac{xe^{i\pi}}{1-x}\right) \nonumber
\end{align}
and
\begin{equation}
\mathcal{F}_{\varepsilon}\left(
\begin{array}
[c]{cc}%
\Delta_{1} & \Delta_{3}\\
\Delta_{2} & \Delta_{4}%
\end{array}
\left|  \Delta_{P}\right|  x\right)  =\sum_{\varepsilon^{\prime}=\text{e,o}%
}\int\frac{dP^{\prime}}{4\pi}K_{\varepsilon}^{\varepsilon^{\prime}}\left(
\begin{array}
[c]{cc}%
\Delta_{1} & \Delta_{3}\\
\Delta_{2} & \Delta_{4}%
\end{array}
|P,P^{\prime}\right)  \mathcal{F}_{\varepsilon^{\prime}}\left(
\begin{array}
[c]{cc}%
\Delta_{1} & \Delta_{3}\\
\Delta_{2} & \Delta_{4}%
\end{array}
\left|  \Delta_{P^{\prime}}\right|  x\right) \label{ker}%
\end{equation}
Here we introduced an index $\varepsilon=\,$e,o to unify the notations, and
$K_{\varepsilon}^{\varepsilon^{\prime}}(P,P^{\prime})$ is the kernel of a
certain integral operator, called the crossing matrix (or, sometimes, the
fusion matrix). In our construction the first relation follow directly from
the symmetries (\ref{Hs}) of the elliptic blocks. The second one is very
non-trivial, the crossing matrix $K_{\varepsilon}^{\varepsilon^{\prime}%
}(P,P^{\prime})$ remaining to be found. In the case of the ordinary
(non-supersymmetric) conformal block an explicit expression for the crossing
matrix has been conjectured in ref.\cite{Ponsot} on the basis of an
appropriate quantum group analysis. Similar analysis seems to be feasible in
the present superconformal situation.

Finally, we hope that the construction of the SLFT four point function,
started in the present paper, will turn useful for the applications in the
Liouville supergravity as well as in non-critical superstring theory.

\vspace{0.4cm}

\textbf{Acknowledgements.} V.Belavin thanks sincerely G.Mussardo and G.Delfino
for their hospitality at SISSA and valuable discussions. His work was
partially supported by the MIUR programme ``Quantum field theory and
statistical mechanics in low dimensions'' and also by the grant RFBR
05-01-01007. A.Belavin acknowledges the hospitality of MPIM during his visit
in February-March 2007 and a support by the grants RFBR 07-02-00799, RFBR-JSPS
05-01-02934, SS-2044-2003 and by the program RAS ''Elementary particles and
Fundamental nuclear physics''. Also he is grateful to R.Poghossian and
L.Hadasz for useful discussions. Both A.Belavin and V.Belavin would like to
express their sincere gratitude to G.Delfino, G.Mussardo, G.von Gehlen,
R.Flume, A.Klumper and T.Miwa, whose help made their collaboration possible at
different stages of the work. Work of Al.Z was sponsored by the European
Committee under contract EUCLID HRPN-CT-2002-00325. An important part has been
made during his visit at the Deptartment of Physics and Astronomy of the
Rutgers University. Hospitality and stimulating scientific atmosphere of the
Theory Group are greatly appreciated, as well as many discussions with A.Zamolodchikov.

\appendix

\section{Superprojective invariance. Three point function}

In this Appendix we consider only Ward identities related to the ``right''
superconformal algebra, formed by the holomorphic components $S(z)$ and $T(z)
$. Respectively, supermultiplets consist of the highest weight vectors $V_{i}$
or $\bar\Lambda_{i}$ defined as in (\ref{W}) (in this section we omit the
parameter $a$ near the primary field to give place for the identification
number $1,2$ etc.) and the ``top components'' $\Lambda_{i}$ or $W_{i}$. To be
definite, we will talk about the multiplet $(V_{i},\Lambda_{i})$. It seems to
require no comments how to combine the results below to the complete
holomorphic-antiholomorphic combinations.

To describe all $2^{3}$ possible three point functions denote
\begin{align}
C_{123} &  =\left\langle V_{1}V_{2}V_{3}\right\rangle \nonumber\\
C_{\widehat{1}23} &  =\left\langle \Lambda_{1}V_{2}V_{3}\right\rangle
\label{C123}\\
&  \text{etc.}\nonumber
\end{align}
From the operator product expansions (\ref{WI}) we have the following
supercurrent Ward identities
\begin{align}
\left\langle S(z)\Lambda_{1}V_{2}V_{3}\right\rangle  &  =\left(  \frac
{2\Delta_{1}}{(z-x_{1})^{2}}+\frac1{z-x_{1}}\frac\partial{\partial x_{1}%
}\right)  C_{123}-\frac1{z-x_{2}}C_{\widehat{1}\widehat{2}3}-\frac1{z-x_{3}%
}C_{\widehat{1}2\widehat{3}}\nonumber\\
\left\langle S(z)V_{1}\Lambda_{2}V_{3}\right\rangle  &  =\left(  \frac
{2\Delta_{2}}{(z-x_{2})^{2}}+\frac1{z-x_{2}}\frac\partial{\partial x_{2}%
}\right)  C_{123}+\frac1{z-x_{1}}C_{\widehat{1}\widehat{2}3}-\frac1{z-x_{3}%
}C_{1\widehat{2}\widehat{3}}\label{Seven}\\
\left\langle S(z)V_{1}V_{2}\Lambda_{3}\right\rangle  &  =\left(  \frac
{2\Delta_{3}}{(z-x_{3})^{2}}+\frac1{z-x_{3}}\frac\partial{\partial x_{3}%
}\right)  C_{123}+\frac1{z-x_{1}}C_{\widehat{1}2\widehat{3}}+\frac1{z-x_{2}%
}C_{1\widehat{2}\widehat{3}}\nonumber
\end{align}
where $\Delta_{1}$, $\Delta_{2}$ and $\Delta_{3}$ are respectively the
dimensions of $V_{1}$, $V_{2}$ and $V_{3}$. As $S(z)=O(z^{-3})$ at
$z\rightarrow\infty$, the following super projective identities hold
\begin{align}
\frac\partial{\partial x_{1}}C_{123} &  =C_{\widehat{1}\widehat{2}%
3}+C_{\widehat{1}2\widehat{3}}\nonumber\\
\frac\partial{\partial x_{2}}C_{123} &  =-C_{\widehat{1}\widehat{2}%
3}+C_{1\widehat{2}\widehat{3}}\nonumber\\
\frac\partial{\partial x_{3}}C_{123} &  =-C_{\widehat{1}2\widehat{3}%
}-C_{1\widehat{2}\widehat{3}}\label{WI1}\\
\left(  x_{1}\frac\partial{\partial x_{1}}+2\Delta_{1}\right)  C_{123} &
=x_{2}C_{\widehat{1}\widehat{2}3}+x_{3}C_{\widehat{1}2\widehat{3}}\nonumber\\
\left(  x_{2}\frac\partial{\partial x_{2}}+2\Delta_{2}\right)  C_{123} &
=-x_{1}C_{\widehat{1}\widehat{2}3}+x_{3}C_{1\widehat{2}\widehat{3}}\nonumber\\
\left(  x_{3}\frac\partial{\partial x_{3}}+2\Delta_{3}\right)  C_{123} &
=-x_{1}C_{\widehat{1}2\widehat{3}}-x_{2}C_{1\widehat{2}\widehat{3}}\nonumber
\end{align}
These identities involve only the correlation functions with even number of
``fermions'' $\Lambda_{i}$. Eliminating the derivatives one finds
\begin{align}
2\Delta_{1}C_{123} &  =-x_{12}C_{\widehat{1}\widehat{2}3}+x_{31}C_{\widehat
{1}2\widehat{3}}\nonumber\\
2\Delta_{2}C_{123} &  =-x_{12}C_{\widehat{1}\widehat{2}3}-x_{23}%
C_{1\widehat{2}\widehat{3}}\label{eqs1}\\
2\Delta_{3}C_{123} &  =x_{31}C_{\widehat{1}2\widehat{3}}-x_{23}C_{1\widehat
{2}\widehat{3}}\nonumber
\end{align}
and thus
\begin{align}
C_{1\widehat{2}\widehat{3}} &  =-\frac{\Delta_{2}+\Delta_{3}-\Delta_{1}%
}{x_{23}}C_{123}\nonumber\\
C_{\widehat{1}2\widehat{3}} &  =\frac{\Delta_{1}+\Delta_{3}-\Delta_{2}}%
{x_{31}}C_{123}\label{CevevC}\\
C_{\widehat{1}\widehat{2}3} &  =-\frac{\Delta_{1}+\Delta_{2}-\Delta_{3}%
}{x_{12}}C_{123}\nonumber
\end{align}
Being substituted to differential equations this sums up to
\begin{align}
\frac\partial{\partial x_{1}}C_{123} &  =-\frac{\Delta_{1}+\Delta_{2}%
-\Delta_{3}}{x_{12}}C_{123}-\frac{\Delta_{1}+\Delta_{3}-\Delta_{2}}{x_{13}%
}C_{123}\nonumber\\
\frac\partial{\partial x_{2}}C_{123} &  =-\frac{\Delta_{1}+\Delta_{2}%
-\Delta_{3}}{x_{21}}C_{123}-\frac{\Delta_{2}+\Delta_{3}-\Delta_{1}}{x_{23}%
}C_{123}\label{diff1}\\
\frac\partial{\partial x_{3}}C_{123} &  =-\frac{\Delta_{1}+\Delta_{3}%
-\Delta_{2}}{x_{31}}C_{123}-\frac{\Delta_{2}+\Delta_{3}-\Delta_{1}}{x_{32}%
}C_{123}\nonumber
\end{align}
and gives finally
\begin{equation}
C_{123}=\frac C{x_{12}^{\Delta_{1}+\Delta_{2}-\Delta_{3}}x_{23}^{\Delta
_{2}+\Delta_{3}-\Delta_{1}}x_{31}^{\Delta_{1}+\Delta_{3}-\Delta_{2}}%
}\label{C123C}%
\end{equation}
where $C$ is an integration constant, independent on $x_{1}$, $x_{2}$ and
$x_{3}$.

Another Ward identity, relevant for the even in fermions functions, is
\begin{align}
\ \left\langle S(z)\Lambda_{1}\Lambda_{2}\Lambda_{3}\right\rangle  &
=\ \left(  \frac{2\Delta_{1}}{(z-x_{1})^{2}}+\frac1{z-x_{1}}\frac
\partial{\partial x_{1}}\right)  C_{1\widehat{2}\widehat{3}}\label{WI2}\\
&  -\left(  \frac{2\Delta_{2}}{(z-x_{2})^{2}}+\frac1{z-x_{2}}\frac
\partial{\partial x_{2}}\right)  C_{\widehat{1}2\widehat{3}}+\left(
\frac{2\Delta_{3}}{(z-x_{3})^{2}}+\frac1{z-x_{3}}\frac\partial{\partial x_{3}%
}\right)  C_{\widehat{1}\widehat{2}3}\nonumber
\end{align}
In the same manner it gives
\begin{align}
\frac\partial{\partial x_{1}}C_{1\widehat{2}\widehat{3}}-\frac\partial
{\partial x_{2}}C_{\widehat{1}2\widehat{3}}+\frac\partial{\partial x_{3}%
}C_{\widehat{1}\widehat{2}3} &  =0\label{deq2}\\
\left(  2\Delta_{1}+x_{1}\frac\partial{\partial x_{1}}\right)  C_{1\widehat
{2}\widehat{3}}-\left(  2\Delta_{2}+x_{2}\frac\partial{\partial x_{2}}\right)
C_{\widehat{1}2\widehat{3}}+\left(  2\Delta_{3}+x_{3}\frac\partial{\partial
x_{3}}\right)  C_{\widehat{1}\widehat{2}3} &  =0\nonumber
\end{align}
It is straightforward to verify that these relations are satisfied identically
with the explicit expressions (\ref{CevevC}).

For the odd fermion number functions consider the Ward identity
\begin{equation}
\left\langle S(z)V_{1}V_{2}V_{3}\right\rangle =\frac1{z-x_{1}}C_{\widehat
{1}23}+\frac1{z-x_{2}}C_{1\widehat{2}3}+\frac1{z-x_{3}}C_{12\widehat{3}%
}\label{WI3}%
\end{equation}
It follows that
\begin{align}
C_{\widehat{1}23}+C_{1\widehat{2}3}+C_{12\widehat{3}} &  =0\label{eq3}\\
x_{1}C_{\widehat{1}23}+x_{2}C_{1\widehat{2}3}+x_{3}C_{12\widehat{3}} &
=0\nonumber
\end{align}
This system is solved in terms of a single function $\tilde C_{123}$
\begin{align*}
C_{\widehat{1}23} &  =x_{23}\tilde C_{123}\\
C_{1\widehat{2}3} &  =x_{31}\tilde C_{123}\\
C_{12\widehat{3}} &  =x_{12}\tilde C_{123}%
\end{align*}
Next, we need the identities
\begin{align}
\left\langle S(z)V_{1}\Lambda_{2}\Lambda_{3}\right\rangle  &  =\frac1{z-x_{1}%
}C_{\widehat{1}\widehat{2}\widehat{3}}\nonumber\\
&  +\left(  \frac{2\Delta_{2}}{(z-x_{2})^{2}}+\frac1{z-x_{2}}\frac
\partial{\partial x_{2}}\right)  C_{12\widehat{3}}-\left(  \frac{2\Delta_{3}%
}{(z-x_{3})^{2}}+\frac1{z-x_{3}}\frac\partial{\partial x_{3}}\right)
C_{1\widehat{2}3}\nonumber\\
\left\langle S(z)\Lambda_{1}V_{2}\Lambda_{3}\right\rangle  &  =-\frac
1{z-x_{2}}C_{\widehat{1}\widehat{2}\widehat{3}}\label{WI4}\\
&  +\left(  \frac{2\Delta_{1}}{(z-x_{1})^{2}}+\frac1{z-x_{1}}\frac
\partial{\partial x_{1}}\right)  C_{12\widehat{3}}-\left(  \frac{2\Delta_{3}%
}{(z-x_{3})^{2}}+\frac1{z-x_{3}}\frac\partial{\partial x_{3}}\right)
C_{\widehat{1}23}\nonumber\\
\left\langle S(z)\Lambda_{1}\Lambda_{2}V_{3}\right\rangle  &  =\frac1{z-x_{3}%
}C_{\widehat{1}\widehat{2}\widehat{3}}\nonumber\\
&  +\left(  \frac{2\Delta_{1}}{(z-x_{1})^{2}}+\frac1{z-x_{1}}\frac
\partial{\partial x_{1}}\right)  C_{1\widehat{2}3}-\left(  \frac{2\Delta_{2}%
}{(z-x_{2})^{2}}+\frac1{z-x_{2}}\frac\partial{\partial x_{2}}\right)
C_{\widehat{1}23}\nonumber
\end{align}
They result in the relations
\begin{align}
\frac\partial{\partial x_{2}}C_{12\widehat{3}}-\frac\partial{\partial x_{3}%
}C_{1\widehat{2}3}+C_{\widehat{1}\widehat{2}\widehat{3}} &  =0\nonumber\\
\frac\partial{\partial x_{1}}C_{12\widehat{3}}-\frac\partial{\partial x_{3}%
}C_{\widehat{1}23}-C_{\widehat{1}\widehat{2}\widehat{3}} &  =0\nonumber\\
\frac\partial{\partial x_{1}}C_{1\widehat{2}3}-\frac\partial{\partial x_{2}%
}C_{\widehat{1}23}+C_{\widehat{1}\widehat{2}\widehat{3}} &  =0\label{eqs4}\\
\left(  2\Delta_{2}+x_{2}\frac\partial{\partial x_{2}}\right)  C_{12\widehat
{3}}-\left(  2\Delta_{3}+x_{3}\frac\partial{\partial x_{3}}\right)
C_{1\widehat{2}3}+x_{1}C_{\widehat{1}\widehat{2}\widehat{3}} &  =0\nonumber\\
\left(  2\Delta_{1}+x_{1}\frac\partial{\partial x_{1}}\right)  C_{12\widehat
{3}}-\left(  2\Delta_{3}+x_{3}\frac\partial{\partial x_{3}}\right)
C_{\widehat{1}23}-x_{2}C_{\widehat{1}\widehat{2}\widehat{3}} &  =0\nonumber\\
\left(  2\Delta_{1}+x_{1}\frac\partial{\partial x_{1}}\right)  C_{1\widehat
{2}3}-\left(  2\Delta_{2}+x_{2}\frac\partial{\partial x_{2}}\right)
C_{\widehat{1}23}+x_{3}C_{\widehat{1}\widehat{2}\widehat{3}} &  =0\nonumber
\end{align}
All of them are satisfied by
\begin{equation}
\tilde C_{123}=\frac{\tilde C}{x_{12}^{\Delta_{1}+\Delta_{2}-\Delta_{3}%
+1/2}x_{23}^{\Delta_{2}+\Delta_{3}-\Delta_{1}+1/2}x_{31}^{\Delta_{1}%
+\Delta_{3}-\Delta_{2}+1/2}}\label{C123tilde}%
\end{equation}
with a new integration constant $\tilde C$, and
\begin{equation}
C_{\widehat{1}\widehat{2}\widehat{3}}=(1/2-\Delta_{1}-\Delta_{2}-\Delta
_{3})\tilde C_{123}\label{C112233}%
\end{equation}

\section{General OPE and special structure constants}

It is instructive to show how the general continuous OPE (\ref{VV}) turns to
the discrete one (\ref{OPE13}) if one of the parameters $a_{1}$ or $a_{2}$ is
set to the degenerate value $-b$. Let us take $a_{2}=-b+\epsilon$ and consider
the first term in (\ref{VV}) with $\mathbb{C}_{a_{1},a_{2}}^{p}$ given by
(\ref{C3})
\begin{align}
\mathbb{C}_{a_{1},a_{2}}^{p}  & =\left(  \pi\mu\gamma\left(  \frac
{Qb}2\right)  b^{1-b^{2}}\right)  ^{(p-a_{1}-a_{2})/b}\times\label{C}\\
& \ \ \ \ \ \frac{\Upsilon_{\text{NS}}^{\prime}(0)\Upsilon_{\text{NS}}%
(2a_{1})\Upsilon_{\text{NS}}(2a_{2})\Upsilon_{\text{NS}}(2Q-2p)}%
{\Upsilon_{\text{NS}}(Q+p-a_{1}-a_{2})\Upsilon_{\text{NS}}(a_{2}%
+p-a_{1})\Upsilon_{\text{NS}}(a_{1}+p-a_{2})\Upsilon_{\text{NS}}(a_{1}%
+a_{2}+p-Q)}\nonumber
\end{align}
At $a_{2}\rightarrow-b$ this expression vanishes due to the zero of the
multiplier $\Upsilon_{\text{NS}}(2a_{2})$. This means that the integral term
in (\ref{VV}) disappears and only the discrete terms contribute. The latter
are due to the singularities of the integral, which come from the pole
structure of the integrand. Expression (\ref{C}) has poles in $p$ at (the four
lines of singularities here correspond respectively to the four multipliers in
the denominator of (\ref{C}))
\begin{align}
&  \ \ \ \ a_{1}+a_{2}-Q-mb^{-1}-nb\;\;\;\text{and\ \ \ \ }a_{1}+a_{2}%
+mb^{-1}+nb\nonumber\\
&  \ \ \ \ a_{1}-a_{2}-mb^{-1}-nb\;\;\;\;\;\;\;\;\text{and\ \ \ \ }%
Q+a_{1}-a_{2}+mb^{-1}+nb\label{polesC}\\
&  \ \ \ \ a_{2}-a_{1}-mb^{-1}-nb\;\;\;\;\;\;\;\;\text{and\ \ \ \ }%
Q+a_{2}-a_{1}+mb^{-1}+nb\nonumber\\
&  \ \ \ \ Q-a_{1}-a_{2}-mb^{-1}-nb\;\;\;\text{and\ \ \ }2Q-a_{1}%
-a_{2}+mb^{-1}+nb\nonumber
\end{align}
where $(m,n)$ -- any pair of non-negative integers of the same parity. At
$a_{2}=-b+\epsilon$ the poles at $p=a_{1}-b+\epsilon$ and $p=a_{1}+b+\epsilon$
of the first multiplier in the denominator of (\ref{C}) come across the poles
at $p=a_{1}-b-\epsilon$ and $p=a_{1}+b-\epsilon$ of the second multiplier
producing two singular terms. The same singularity appears from the two
``reflection symmetric'' pinches at $p=Q-a_{1}+b$ and $p=Q-a_{1}-b$. Due to
the symmetry properties of the integrand and the reflection relation
(\ref{reflection}) the ``reflected'' terms give the same contributions and
don't need separate consideration.

First, let us pick up the pole at $p=a_{1}-b-\epsilon$%
\begin{equation}
\operatorname*{res}_{p=a_{1}-b-\epsilon}\mathbb{C}_{a_{1},a_{2}}^{p}%
=\frac{\Upsilon_{\text{NS}}^{\prime}(0)\Upsilon_{\text{NS}}(2a_{1}%
)\Upsilon_{\text{NS}}(-2b+2\epsilon)\Upsilon_{\text{NS}}(2Q-2a_{1}%
+2b)}{\Upsilon_{\text{NS}}(Q-2\epsilon)\Upsilon_{\text{NS}}^{\prime
}(-2b)\Upsilon_{\text{NS}}(2a_{1})\Upsilon_{\text{NS}}(2a_{1}-2b-Q)}%
=1\label{resCm}%
\end{equation}
Similar calculation for the residue at $p=a_{1}+b-\epsilon$ results in
\begin{align}
\operatorname*{res}_{p=a_{1}+b-\epsilon}\mathbb{C}_{a_{1},a_{2}}^{p} &
=\left(  \pi\mu\gamma\left(  \frac{Qb}2\right)  b^{1-b^{2}}\right)
^{2}\ \frac{\Upsilon_{\text{NS}}(2a_{1})\Upsilon_{\text{NS}}(2a_{1}%
+2b-Q)}{\Upsilon_{\text{NS}}(2a_{1}+2b)\Upsilon_{\text{NS}}(2a_{1}%
-Q)}\label{resCp}\\
&  =\left(  \gamma\left(  \frac{Qb}2\right)  \right)  ^{2}\ \frac{(\pi\mu
)^{2}b^{4}\gamma(a_{1}b-1/2-b^{2}/2)}{\gamma(1/2+b^{2}/2+a_{1}b)}\nonumber
\end{align}
where we systematically use the shift relations (\ref{Yshift}).

Second, let's work out the contribution of the second term with
\begin{align}
\mathbb{\tilde C}_{a_{1},a_{2}}^{p} &  =\left(  \pi\mu\gamma\left(  \frac
{Qb}2\right)  b^{1-b^{2}}\right)  ^{(p-a_{1}-a_{2})/b}\times\label{Ctilde}\\
&  \ \ \ \ \ \ \frac{2i\Upsilon_{\text{NS}}^{\prime}(0)\Upsilon_{\text{NS}%
}(2a_{1})\Upsilon_{\text{NS}}(2a_{2})\Upsilon_{\text{NS}}(2Q-2p)}%
{\Upsilon_{\text{R}}(Q+p-a_{1}-a_{2})\Upsilon_{\text{R}}(a_{2}+p-a_{1}%
)\Upsilon_{\text{R}}(a_{1}+p-a_{2})\Upsilon_{\text{R}}(a_{1}+a_{2}%
+p-Q)}\nonumber
\end{align}
The pole structure is given by the same formula (\ref{polesC}) where now
$(m,n)$ is a pair of non-negative integers of opposite parity. At
$a_{2}=-b+\epsilon$ we have to pick up a singular term at $p=a_{1}-\epsilon$
\begin{align}
\operatorname*{res}_{p=a_{1}-\epsilon}\mathbb{\tilde C}_{a_{1},-b+\epsilon
}^{p} &  =\left(  \pi\mu\gamma\left(  \frac{Qb}2\right)  \right)
\frac{2i\Upsilon_{\text{NS}}^{\prime}(0)\Upsilon_{\text{NS}}(2a_{1}%
)\Upsilon_{\text{NS}}(-2b+2\epsilon)\Upsilon_{\text{NS}}(2Q-2p)}%
{\Upsilon_{\text{R}}(Q+b-2\epsilon)\Upsilon_{\text{R}}^{\prime}(-b)\Upsilon
_{\text{R}}(2a_{1}+b)\Upsilon_{\text{R}}(2a_{1}-b-Q)}\label{resC0}\\
&  =\frac{2\pi i\mu}{\gamma\left(  -b^{2}\right)  \gamma\left(  ba_{1}\right)
\gamma\left(  1+b^{2}-ba_{1}\right)  }\nonumber
\end{align}
Residues (\ref{resCm}), (\ref{resCp}) and (\ref{resC0}) can be compared with
the special stricture constants, derived in section 2 in terms of the
``screening'' integrals.

\section{Dotsenko-Fateev type equation}

Substitution
\begin{equation}
g=x^{a_{1}b}(1-x)^{a_{2}b}F\label{gF}%
\end{equation}
renders eq.(\ref{diffeq}) to the form
\begin{align}
x^{2}(1-x)^{2}F^{\prime\prime\prime}-x(1-x)(K_{1}x-K_{2}(1-x))F^{\prime
\prime}  & +(L_{1}x^{2}+L_{2}(1-x)^{2}-L_{3}x(1-x))F^{\prime}\label{FDeq}\\
\  & +(M_{1}x-M_{2}(1-x))F=0\nonumber
\end{align}
where
\begin{align}
K_{1}  & =-2g-3B-3C\,;\;\;\;\;\;\;\;\;\;\;\;\;\;\;\;\;\;\;\;\;\;\;K_{2}%
=-2g-3A-3C\nonumber\\
L_{1}  & =(B+C)(2B+2C+2g+1)\,;\;\;\;\;L_{2}=(A+C)(2A+2C+2g+1)\nonumber\\
L_{3}  & =4AB+4(2A+2B+2C+1)C+4(A+B+3C)g+\allowbreak4g^{2}+2g\label{KLM}\\
M_{1}  & =-2C(A+B+C+g+1)(2B+2C+2g+1)\,\nonumber\\
M_{2}  & =-2C(A+B+C+g+1)(2A+2C+2g+1)\nonumber
\end{align}
while the parameters $A$, $B$, $C$ and $g$ are related to $a_{1}$, $a_{2}$,
$a_{3}$ and $b$ as in eq.(\ref{ABC}).

Consider the two-fold contour integrals
\begin{equation}
I_{\alpha\beta}(x)=%
{\displaystyle\int\limits_{C_{\alpha}}}
{\displaystyle\int\limits_{C_{\beta}}}
dt_{1}dt_{2}\left|  t_{1}t_{2}\right|  ^{A}\left|  (1-t_{1})(1-t_{2})\right|
^{B}\left|  (x-t_{1})(x-t_{2})\right|  ^{C}\left|  t_{1}-t_{2}\right|
^{2g}\label{Iab}%
\end{equation}
at $\alpha\neq\beta$ and
\begin{equation}
I_{\alpha\alpha}(x)=\frac12%
{\displaystyle\int\limits_{C_{\alpha}}}
{\displaystyle\int\limits_{C_{a}}}
dt_{1}dt_{2}\left|  t_{1}t_{2}\right|  ^{A}\left|  (1-t_{1})(1-t_{2})\right|
^{B}\left|  (x-t_{1})(x-t_{2})\right|  ^{C}\left|  t_{1}-t_{2}\right|
^{2g}\label{Iaa}%
\end{equation}
where the contours $C_{\alpha}$, $\alpha=1,2,3,4$ are numbered as follows
\begin{align}
C_{1}=(-\infty,0]\,;\;\;C_{2}=[0,x]\,;\;\;C_{3}=[x,1]\,;\;\;C_{4}%
=[1,\infty)\label{Ca}%
\end{align}
It is verified directly\footnote{It is implied that the parameters are chosen
in a way to ensure convergence of all these integrals. Otherwise, standard
regularization is in order.} that all these integrals are solutions to
eq.(\ref{FDeq}).

Of all the nine integrals only three are independent. As a base one can choose
the set with a diagonal monodromy around the point $x=0$%
\begin{align}
\mathcal{I}_{1}(x)  & =I_{44}(x)\sim\mathcal{I}_{1}^{(0)}(1+\ldots)\nonumber\\
\mathcal{I}_{2}(x)  & =I_{24}(x)\sim x^{1+A+C}\mathcal{I}_{2}^{(0)}%
(1+\ldots)\label{I123}\\
\mathcal{I}_{3}(x)  & =I_{22}(x)\sim x^{2+2A+2C+2g}\mathcal{I}_{3}%
^{(0)}(1+\ldots)\nonumber
\end{align}
where $\ldots$ stands for a regular series in $x$. The constants
\begin{align}
\mathcal{I}_{1}^{(0)}  & =\frac{\Gamma(2g)\Gamma(1+B)\Gamma(1+B+g)\Gamma
(-1-2g-A-B-C)\Gamma(-1-g-A-B-C)}{\Gamma(g)\Gamma(-g-A-C)\Gamma(-A-C)}%
\nonumber\\
\mathcal{I}_{2}^{(0)}  & =\frac{\Gamma(1+A)\Gamma(1+B)\Gamma(1+C)\Gamma
(-1-2g-A-B-C)}{\Gamma(2+A+C)\Gamma(-2g-A-C)}\label{Iconst}\\
\mathcal{I}_{3}^{(0)}  & =\frac{\Gamma(2g)\Gamma(1+A)\Gamma(1+A+g)\Gamma
(1+C)\Gamma(1+C+g)}{\Gamma(g)\Gamma(2+A+C+g)\Gamma(2+A+C+2g)}\nonumber
\end{align}
are calculated using the Selberg integral\cite{Selberg}
\begin{equation}
\frac1{n!}%
{\displaystyle\int\limits_{0}^{1}}
{\displaystyle\prod\limits_{i=1}^{n}}
dt_{i}t_{i}^{\mu-1}(1-t_{i})^{\nu-1}\prod\limits_{i>j}\left|  t_{i}%
-t_{j}\right|  ^{2g}=%
{\displaystyle\prod\limits_{k=0}^{n-1}}
\frac{\Gamma(g+kg)\Gamma(\mu+kg)\Gamma(\nu+kg)}{\Gamma(g)\Gamma(\mu
+\nu+(n-1+k)g)}\label{Selberg}%
\end{equation}
Another base
\begin{align}
\mathcal{J}_{1}(x)  & =I_{11}(x)\sim\mathcal{J}_{1}^{(0)}(1+\ldots)\nonumber\\
\mathcal{J}_{2}(x)  & =I_{13}(x)\sim(1-x)^{1+B+C}\mathcal{J}_{2}%
^{(0)}(1+\ldots)\label{J123}\\
\mathcal{J}_{3}(x)  & =I_{33}(x)\sim(1-x)^{2+2B+2C+2g}\mathcal{J}_{3}%
^{(0)}(1+\ldots)\nonumber
\end{align}
where the dots now replace a regular series in $1-x$, enjoys diagonal
monodromy around $x=1$. Apparently%

\begin{equation}
\mathcal{I}_{\alpha}=\sum_{\beta}\mathcal{M}_{\alpha\beta}\mathcal{J}_{\beta
}\label{IJ}%
\end{equation}
where the ``fusion matrix'' $\mathcal{M}_{\alpha\beta}$ is evaluated by
manipulating the contours of integration. It has the following entries
\cite{FD2}
\begin{align}
\mathcal{M}_{11}  & =\dfrac{\sin\pi A\sin\pi(g+A)}{\sin\pi(C+B)\sin\pi
(g+C+B)}\nonumber\\
\mathcal{M}_{12}  & =-\dfrac{\sin\pi A\sin\pi C}{\sin\pi(C+B)\sin\pi
(2g+C+B)}\nonumber\\
\mathcal{M}_{13}  & =\dfrac{\sin\pi C\sin\pi(g+C)}{\sin\pi(g+C+B)\sin
\pi(2g+C+B)}\nonumber\\
\mathcal{M}_{21}  & =-\dfrac{2\cos\pi g\sin\pi(g+A)\sin\pi(g+A+B+C)}{\sin
\pi(C+B)\sin\pi(g+C+B)}\label{Mab}\\
\mathcal{M}_{22}  & =\dfrac{\sin\pi C\sin\pi(g+A+B+C)}{\sin\pi(C+B)\sin
\pi(g+C+B)}-\dfrac{\sin\pi(g+B)\sin\pi A}{\sin\pi(g+C+B)\sin\pi(2g+C+B)}%
\nonumber\\
\mathcal{M}_{23}  & =\dfrac{2\cos\pi g\sin\pi(g+C)\sin\pi(g+B)}{\sin
\pi(g+C+B)\sin\pi(2g+C+B)}\nonumber\\
\mathcal{M}_{31}  & =\dfrac{\sin\pi(g+A+B+C)\sin\pi(2g+A+B+C)}{\sin
\pi(C+B)\sin\pi(g+C+B)}\nonumber\\
\mathcal{M}_{32}  & =\dfrac{\sin\pi B\sin\pi(2g+A+B+C)}{\sin\pi(C+B)\sin
\pi(2g+C+B)}\nonumber\\
\mathcal{M}_{33}  & =\dfrac{\sin\pi B\sin\pi(g+B)}{\sin\pi(g+C+B)\sin
\pi(2g+C+B)}\nonumber
\end{align}
Next, it is verified\cite{FD2} that the combination
\begin{equation}
X_{1}\mathcal{I}_{1}(x)\mathcal{I}_{1}(\bar x)+X_{2}\mathcal{I}_{2}%
(x)\mathcal{I}_{2}(\bar x)+X_{3}\mathcal{I}_{3}(x)\mathcal{I}_{3}(\bar
x)\label{Xc}%
\end{equation}
is a single-valued function of $(x,\bar x)$ if
\begin{align}
\frac{X_{3}}{X_{1}}  & =\frac{\sin\pi A\sin\pi C\sin\pi(A+C)\sin\pi
(A+g)\sin\pi(C+g)}{\sin\pi B\sin\pi(B+g)\sin\pi(A+B+C+g)\sin\pi(A+C+2g)\sin
\pi(A+B+C+2g)}\nonumber\\
\frac{X_{2}}{X_{1}}  & =\frac{\sin\pi(A+C+g)\sin\pi A\sin\pi C}{2\cos\pi
g\sin\pi(B+g)\sin\pi(A+B+C+g)\sin\pi(A+C+2g)}\label{XX}%
\end{align}

\end{document}